\documentclass[12pt,oneside]{iitmMyPhD}
\usepackage{psfig, epsf, cite, float, subfigure, graphicx, verbatim,amsmath,amsbsy,txfonts,multirow,lscape,amsfonts}
\usepackage{array}
\usepackage{rotating}
\usepackage{appendix}
\usepackage[breaklinks=true]{hyperref}
\renewcommand{\baselinestretch}{1} 

\def \ma {\mathcal{A}}
\def \mi {\mathcal{I}}
\def \ml {\mathcal{L}}
\def \mr {\mathcal{R}}
\def \mb {\mathcal{B}}
\def \ms {\mathcal{S}}

\def \mc {\mathcal{C}}
\def \mt {\mathcal{T}}
\def \mg {\mathcal{G}}
\def \mk {\mathcal{K}}
\def \mi {\mathcal{I}}
\def \mj {\mathcal{J}}
\def \mj {\mathcal{Q}}
\def \mn {\mathcal{N}}

\def \ck {\mathcal{K}}
\def \cn {\mathcal{N}}
\def \ct {\mathcal{T}}
\def \cs {\mathcal{S}}
\def \cg {\mathcal{G}}
\def \calr {\mathcal{R}}

\def \mq {\mathbb{Q}}
\def \mm {\mathfrak{M}}

\def \alc {\mathcal{ALC}}

\def \shoiq {\mathcal{SHOIQ}}
\def \sroiq {\mathcal{SROIQ}}
\def \sroiqc {\mathcal{SROIQ(C)}}
\def \gcsr {\mathcal{GC}$-$\mathcal{SROIQ(C)}}

\def \vat {V_a^T}
\def \vct {V_c^T}
\def \eat {\mathcal{E}_a^T}
\def \ect {\mathcal{E}_c^T}
\def \lt {\mathcal{L}^T}
\def \lst {\mathcal{L}^T(s)}
\def \ltt {\mathcal{L}^T(t)}

\def \vas {V_a^S}
\def \vcs {V_c^S}
\def \eas {E_a^S}
\def \ecs {E_c^S}
\def \cts {\mathcal{Q}}
\def \mss {M^S}
\def \ls {\mathcal{L}^S}
\def \ls {\mathcal{L}(s)}
\def \lx {\mathcal{L}(x)}

\def \ly {\mathcal{L}(y)}
\def \ls {\mathcal{L}^S}

\def \la {\langle}
\def \ra {\rangle}

\def \rar {\rightarrow}

\def \sqeq {\sqsubseteq}
\def \sqeqs {\stackrel{*}{\sqeq}}

\def \dmi {\Delta^\mi}

\usepackage{epstopdf}
\usepackage[linesnumbered,ruled,vlined]{algorithm2e}

\begin{document}


\thispagestyle{empty}
\setcounter{page}{0}
\vglue 0in
\begin{center}
	{\Large{\bf Extending SROIQ with Constraint Networks and Grounded Circumscription  }}\\[10ex]
	{\em MTP Report submitted to}\\
	{\em Indian Institute of Technology, Mandi}\\
	{\em for the award of the degree}\\[2ex]
	{\em of}\\[5ex]
	{\large \bf  B. Tech} \\ [2ex]
	{\normalsize \em by}\\[5ex]
	{\large{\bf Arjun Bhardwaj}}\\[2ex]
	{\em under the guidance of}\\[2ex]
	{\bf Prof. Deepak Khemani (Guide)}\\
	{\bf Prof. Dileep A.D. (Co-Guide)}
\end{center}

\vglue 2ex plus 1fill  
\vglue 4ex
\centerline{\includegraphics[width=4.5cm]{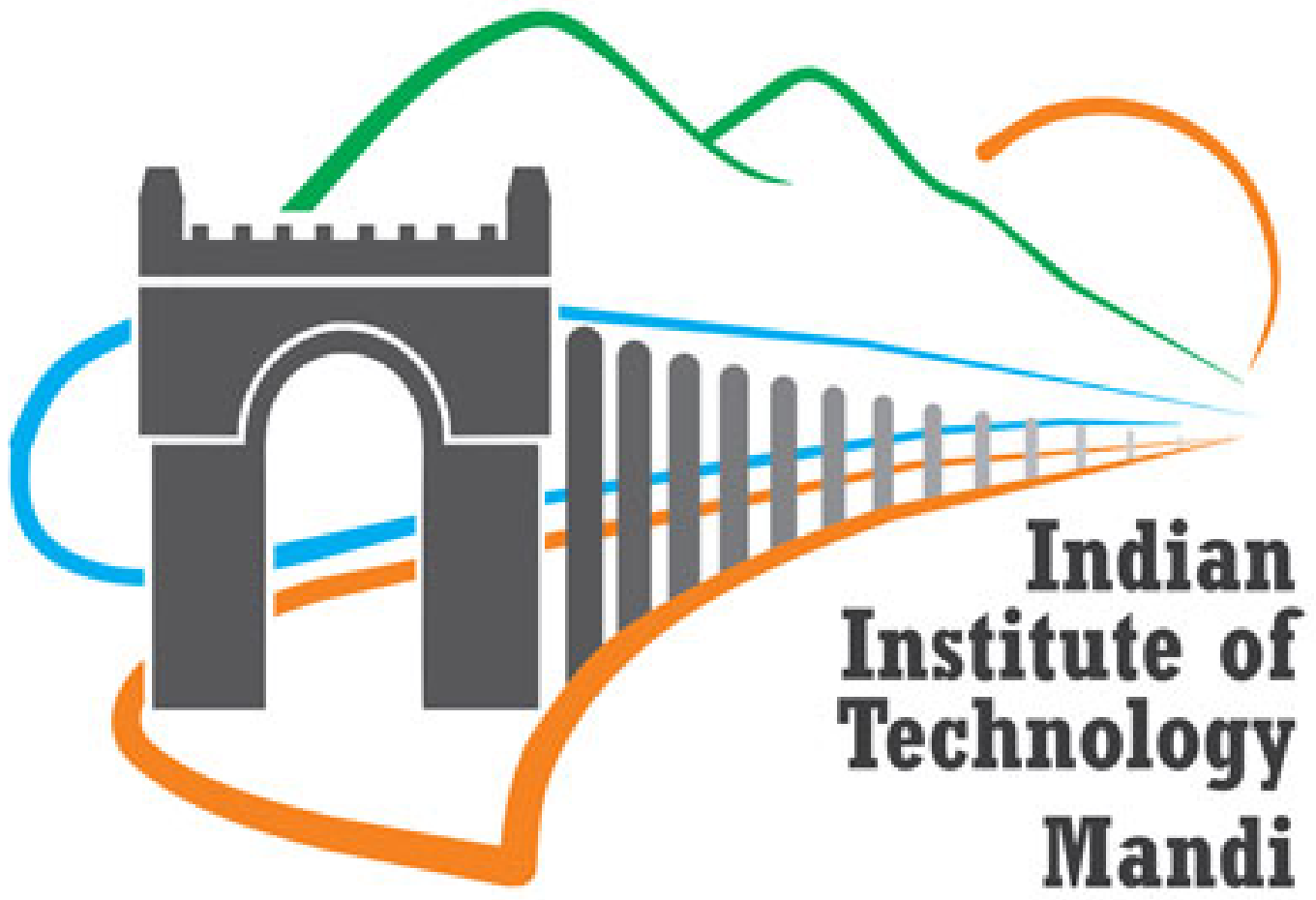}}

\begin{center}
	{\normalsize \bf SCHOOL OF COMPUTING AND ELECTRICAL ENGINEERING}\\[1ex]
	{\large \bf INDIAN INSTITUTE OF TECHNOLOGY, MANDI}\\[1ex]
	{\bf JUNE 2015}
\end{center}
\clearpage
\thispagestyle{empty}

 \pagestyle{plain}					
  \addcontentsline{toc}{chapter}{Thesis Certificate}
        \pagenumbering{roman} \setcounter{page}{1}

\chapter*{Acknowledgments}

I thank Prof. Deepak Khemani for his kind support and guidance, throughout the course of the project.
I thank Prof. Dileep A. D. for his reviews of the work.

The problem dealt with, in the project, originated out of discussions with Sangeetha, AIDB Lab, IIT Madras.
I thank her immensely for her expertise, helpful criticism and for her contributions towards framing and formalizing the research problem.

~\hfill
{\it Arjun Bhardwaj}

\clearpage 
      
\addcontentsline{toc}{chapter}{Abstract}		
\chapter*{Abstract}
\label{abstract}

Developments in semantic web technologies have promoted ontological encoding of knowledge from diverse domains. 
However, modelling many practical domains requires more expressiveness than what the standard description logics (most prominently SROIQ) support.
In this paper, we extend the expressive DL SROIQ with constraint networks (resulting in the logic SROIQc) and grounded circumscription (resulting in the logic GC-SROIQ).
Applications of constraint modelling include embedding ontologies with temporal or spatial information, while those of grounded circumscription include defeasible inference and closed world reasoning.

We describe the syntax and semantics of the logic formed by including constraint modelling constructs in SROIQ, and provide a sound, complete and terminating tableau algorithm for it.
We further provide an intuitive algorithm for Grounded Circumscription in SROIQ, which adheres to the general framework of grounded circumscription, and which can be applied to a whole range of expressive logics for which no such specific algorithm presently exists.\\\\

{\bf Keywords:} {\em SROIQ, Constraint Networks, Grounded Circumscription, Description Logic, Tableau}			

\tableofcontents        			


\pagenumbering{arabic} \setcounter{page}{1}
\chapter{Introduction}
\label{chapter1}

The issue of knowledge representation lies at the heart of the quest for artificial intelligence.
So, when First Order Logic (FOL) was formalized, at first it seemed like everything was solved.
But it was soon observed that the semi-decidability of FOL and high time complexity of reasoning in FOL prohibited its use for reasoning over real-world-scale Knowledge Bases.
This points us to the general phenomenon known as the representation-reasoning trade-off.
If the logic used for representing knowledge is very expressive, and thus can be used for representing more complex pieces of information, then reasoning over such a logic would also be more difficult.

Description Logics (DLs) represent a comfortable compromise between representational expressiveness and ease of reasoning.
They are a popular choices for modelling domain knowledge, given that inference procedures in these logics is decidable, sound and complete.
The developments in DLs have directly fuelled semantic web technologies which depend on them to provide reasoning support over real world ontologies.

Recent years have seen a trend towards formalizing and capturing knowledge from diverse domains in the form of ontologies.
Standard DLs are able to provide reasoning support for the vast majority of ontologies.
However, there are some domains which require more expressive representation mechanisms to formalize their knowledge.
We focus on two such popularly proposed extensions.

Several ontologies require the ability to embed spacial and temporal information about the entities being modelled.
Further, many ontologies require the general constraint modelling ability to formally specify the constraints between entities in terms of a constraint network.
An important result, provided by \cite{CDR}, allows us to merge the above two.
Constraint networks have been proposed as a viable way to capture temporal and spacial information.
Most alternative techniques compromise on either expressiveness of constraint modelling or come at the cost of heightened time complexity of the resulting logic.
The importance of embedding temporal information is discussed in \cite{towl}, which also provides a guideline on the constructs that must be made available to OWL to allow expressive modelling of temporal information.

On the other hand, DLs are all open domain.
Intuitively this means that what is expressed in the Knowledge Base is true, but one can not assume that what is not mentioned to be false.
This poses difficulties in reasoning in many cases, and hence several ways have been explored to facilitate closed world reasoning in DLs, which is in accordance with the assumption that only what is expressed is true.
Additionally, many domains have rules which apply "in general", but not always.
The strict logical rules offered by DLs are not able to correctly model these default rules.
Out of the several solutions that have been proposed to the meet the above requirements, grounded circumscription \cite{lcws} is a formalism that provides an intuitive and elegant solution.

In this paper, we extend the expressive DL $\sroiq$ \cite{sroiq} with constraint networks and grounded circumscription.
The aim is to formalize the logic by defining its syntax and semantics and to propose a tableau decision procedure to support reasoning in it.

\section{Related Work}

Earlier attempts to embed temporal information into DLs have only been realized for less expressive logics like $\alc$. 
For example \cite{tdl}.
The work by \cite{CDR} augments $\alc$ with $\omega$-admissible Constraint Systems to form a decidable DL. 
The $\omega$-admissible systems identified were the Allens' relations (for temporal intervals) and RCC8 relations (for spatial regions).

Circumscription is a non-monotonic framework that allows closed world reasoning and default reasoning. 
However, minimization of roles is undecidable even for simpler DLs and tableau procedures exist only for concept minimization in logics with finite model property. 
\cite{lcws} uses grounded circumscription, using which both concept and role minimization in DLs is decidable. However, a tableau procedure exists only for ALCO and involves special rules to ensure the construction of a grounded model.

\section{Contribution of the work}

The current work incorporates into $\sroiq$, the constraint modelling constructs, including those mentioned in \cite{CDR}.
We also specify the conditions which must be imposed on these constructs to ensure the existence of a tableau algorithm for the resulting logic.
Besides these constructs we introduce the following features for constraint modelling :
\begin{itemize}
\item
ability to explicitly name variables in constraint networks
\item
simple hierarchies in concrete roles
\item
constraint modelling constructs that make use of the named variables
\item
number restrictions on concrete roles.
\end{itemize} 
Also provided are sketches of the proofs of correctness of the resulting tableau algorithm.\\

We further provide the details of a intuitive, general algorithm that can be used to allow circumscription circumscription in $\sroiq$ and a range of other logics.
Unlike existing algorithms, it simply uses some of the constructs of $\sroiq$, to perform the tasks outlined in the grounded circumscription framework \cite{lcws}.

\section{Organization of the Report}

In chapter 2, we explore the extension of $\sroiq$ to include constraint networks.
We provide the syntax, semantics and the tableau inference procedure of the resulting logic.
We argue for its correctness by providing sketches of corresponding proofs in the Appendix.
We argue that more expressive constraint modelling constructs pose several fundamental, unavoidable difficulties towards the design of a tableau algorithm for the logic.

In the following chapter we take up the case of grounded circumscription and provide an algorithm that permits grounded circumscription in $\sroiq$.


\chapter{Augmenting SROIQ with Constraint Networks}
\label{chapter3}

Here we present the logic that provides constraint modelling constructs to $\sroiq$.


\section{Constraint Systems}

We use the notion of a constraint system as defined in \cite{CDR}. Let \textit{Var} be a countably infinite set of variables and \textit{Rel} a finite set of relation symbols.
While describing the algorithm, we may use the term concrete domain to denote the set \textit{Var}.

A \textbf{\textit{Rel}-constraint} is an expression of the form $(v\:r\:v')$, where $v,v'\in$\textit{Var}, $r\in$\textit{Rel}. 
A \textbf{\textit{Rel}-network} is a finite set of \textit{Rel}-constraints. For a \textit{Rel}-network $N$, let $V_N$ denote the variables used in the network. 

$N$ is \textbf{complete} if for every $v,v'\in V_N$, there is exactly one constraint $(v\:r\:v')\in N$. 
Let \textbf{$\mm$} be the set of all complete networks, possible according to the semantics of the domain being modelled. 
A network $N'$ is a \textbf{model} of a network $N$, if $N'\in\mm$ and there is a mapping $\tau: V_{N} \rar V_N'$ such that $(v\:r\:v')\in N$ implies $(\tau(v)\:r\:\tau(v'))\in N'$.
A network $N$ is \textbf{satisfiable} in $\mc$ if $\mm$ contains a model of $N$.
We define a \textbf{constraint system} $\mc$ as the tuple $\la Rel,\mm \ra$.

Next, we define the some terms that help identify the set of constraint systems which can be used with the algorithm, without the loss of decidability.

Given any finite constraint networks $N,M$, their \textbf{intersection} is defined as follows:\\
$I_{N,M} := \{(v\:r\:v')\mid v,v'\in V_N\cap V_M \text{ and }(v\:r\:v')\in N\}$

$\mc$ has the \textbf{patchwork property} if for any finite and satisfiable networks $M,N$ with complete and equal intersections (i.e. $I_{M,N}$ and $I_{N,M}$ are complete networks, $I_{M,N}=I_{N,M}$), $N\cup M$ is satisfiable.
$\mc$ has the \textbf{compactness property} if the following holds: a network $N$ with infinite set of variables ($V_N$) is satisfiable, if and only if, for every finite $V\subseteq V_N$, the network $N\mid _V\:=\:\{(v\:r\:v')\in N\:\mid\:v,v'\in V\}$ is satisfiable.
We say that $\mc$ is \textbf{$\omega$-admissible} if the following holds
\begin{enumerate}
\item Satisfiability in $\mc$ is decidable
\item $\mc$ has patchwork property
\item $\mc$ has the compactness property
\end{enumerate}
To ensure decidability, only $\omega$-admissible constraint system are permitted to be used in tableau algorithms for constraint modelling logics described here.


\section{The Logic $\sroiqc$}

We describe $\sroiqc$ : a logic that provides relatively controlled constraint modelling constructs on top of $\sroiq$ \cite{sroiq}, in order to ensure the existence of a tableau for inference over the language.
In this section, we describe the constructs of the logic, their semantics and provide a tableau decision procedure for $\sroiqc$-KB satisfiability.


\subsection{Syntax}
Let $N_C$ be the set of concept names, $Roles$ the set of abstract role names (including the universal role $U$), $N_{cR}$ the set of concrete role names, $N_{aI}$ names of abstract individuals, $Nom$ the set of nominal concept names (also called nominals), $Var$ the set of variables which participate in the $Rel$-constraints and $N_{cI}\subset Var$ the set of constraint individual names.

A $\sroiqc$ Knowledge Base ($\mk$) consists of a tuple $\la\ma,\mt,\mr\ra$ where $\ma,\mt,\mr$ are respectively the A-Box, T-Box and R-Box, which are described as follows. Since the logic is built on top of $\sroiq$, it shares most of its constructs with it.
We include description of these here for completeness, and the user may refer to \cite{sroiq} for more detailed treatment.

\paragraph*{ R-Box}
$\mr$ is defined as the tuple $\la\:\mr_h,\mr_{ch},\mr_a\:\ra$.
We define these, in the subsequent paragraphs.

the \textbf{Inverse} of a role is defined as follows :
\begin{equation}
Inv(R) \: =
\begin{cases}
R^- & \text{ if $R\in Roles$}
\\
S & \text{ if $R=S^-$ for some $S\in Roles$}
\end{cases}
\end{equation}
We define $N_{aR}=Roles\cup\{R^{-}\:\mid\:R\in Roles\}$

A \textbf{role chain} is an expression of the form $R_1 \dots R_n$ with $n\geq 1$ and each $R_i\in N_{aR}$.
The notion of $Inv$ is extended to role chains as : 
\[Inv(R_1R_2\dots R_n)=Inv(R_n)\dots Inv(R_2)Inv(R_1)\]

$\mr_{ch}$ is the set of concrete Role Inclusion Axioms (c-RIA) each of which is of the form $g\sqeq g'$ where $g,g'\in N_{cR}$. This defines a hierarchy between the concrete roles.
$\mr_{ah}$ is a finite set of abstract Role Inclusion Axioms (a-RIA) of the form $w\sqeq R$ where $w$ is a role chain, possibly of unit length and $R\in N_{aR}$.
However, in order to ensure decidability of the tableau algorithm, the set $\mr_{ah}$ must be a Regular. The notion is explained below.
$\sqeqs$ is the closure of $\sqeq$ relation between roles.

A \textbf{regular order} $\prec$ is an irreflexive transitive binary relation on the set of roles $N_{aR}$ satisfying : $R_1 \prec R_2$ iff $Inv(R_1)\prec Inv(R_2)$.
$\mr_{ah}$ is \textbf{Regular} if there is a regular order on $N_{aR}$ such that each RIA in $\mr$ is one of the following forms:
\begin{enumerate}
\item $R_1\dots R_n\sqeq R$ with $R_i\prec R$ for all $i\leq i \leq n$
\item $RR_1\dots R_n\sqeq R$ with $R_i\prec R$ for all $i\leq i \leq n$
\item $R_1\dots R_nR\sqeq R$ with v$R_i\prec R$ for all $i\leq i \leq n$
\item $RR\sqeq R$
\item $Inv(R)\sqeq R$
\end{enumerate}

Associated with the notion of regularity of an abstract is the notion of a simple role. 
Given a role hierarchy $\mr_{ah}$, the set of roles that are \textbf{simple} in $N_{aR}$, is inductively defined as follows:
\begin{itemize}
\item A role name is simple if it does not occur on the R.H.S. of a RIA in $\mr_h$
\item An inverse role $R^-$ is simple if $R$ is, and
\item If $R$ occurs on the R.H.S. of a RIA in $\mr_h$, then $R$ is simple if, for each $w\sqeq R \in \mr_h\:,\: w=S$ for a simple role $S$
\end{itemize}
We further exclude the universal role $U$ from being in the set of simple roles. An RIA is $simple$ if it is of the form $S\sqsubseteq S'$, where $S,S'$ are simple roles.

A \textit{path} is a sequence $R_1,\dots,R_ng$ consisting of simple roles $R_1\dots, R_k\in N_{aR}$ and a concrete role $g\in N_{cR}$.
We define the length of a path as the number of abstract or concrete roles that appear in the sequence.

$\mr_a$ is a finite set of \textit{role assertions}.
A role assertion is of the form 
$Ref(R)$ (reflexivity),
$Irr(S)$ (irreflexivity), 
$Dis(S,S')$ (role disjointness), 
$Sym(R)$ (symmetry),
$Trans(R)$ (transitivity),
$Fxnl(X)$ (functionality), where $S,R\in N_{aR}$, $S$ is a simple role and $X\in N_{aR}\cup N_{cR}$ must be a simple role if it is an abstract role.

In conclusion, the R-Box features novel to $\sroiqc$ w.r.t $\sroiq$ include the concrete role hierarchy ($\mr_{ch}$), and the functional role assertion for concrete roles.

\paragraph*{Notational Conventions} Unless mentioned otherwise, the following notational conventions are followed : $r,r'\in Rel(\mc)$; $g,g'\in N_{cR}$; $R,R'\in N_{aR}$; $S,S'\in N_{aR}$ are simple abstract roles; $A,B,C,D$ are concepts; $o\in Nom$ i.e. $o$ is a nominal set; $i\in N_{cI}$ i.e. $i$ is a named variable - a constraint individual, $a,b\in N_{aI}$ i.e. they are abstract named individuals from the domain being modelled; $U$ is a path; $G,G'$ are paths. 
The same applies to respective symbols with subscripts.

\paragraph{The T-Box}
The set of $\sroiqc$ concepts, $Concepts$, is defined recursively as follows:
\begin{align*}
& A \mid o \mid (C\sqcap D) \mid (C\sqcup D) \mid \neg C \mid \exists R.C \mid \forall R.C \mid \\
&\geq n S.C \mid \leq n S.C \mid \exists S.Self \mid\\
&\leq_c ng  \mid \geq_c ng \mid \\ &
\exists_cU_1,U_2.r \mid \exists_cU_1,\{i\}.r \mid \exists_c\{i\},U_1.r \mid \\
&\forall_cU_1,U_2.r \mid \forall_cU_1,\{i\}.r \mid \forall_c\{i\},U_1.r 
\end{align*}
Here $A\in N_C$, $C,D\in Concepts$.
In the above, all the non-$\sroiq$ concepts (the last three lines) are known as concrete concepts.

There are some constraints imposed on the paths that comprise the concrete concepts.
For concepts of the form $(\exists_cU_1,\{i\}.r), (\exists_c\{i\},U_1.r), (\forall_cU_1,\{i\}.r), (\forall_c\{i\},U_1.r)$, $U_1$ is restricted to be a path of length at most 2 i.e. it can be of the form $Rg$ or $g$.
For the concepts $(\exists_c U_1,U_2.r)$ and $(\forall_c U_1,U_2.r)$, if one of the paths is of length 2, then the other must be of the length 1.
This makes the permissible combinations of $(U_1,U_2)$ to be of the form $(Rg,g)$, or $(g,g)$ or $(g,Rg)$.
This condition is the Path Normal Form, as defined in \cite{CDR}.

A $\sroiqc$ TBox contains GCIs of the form $C\sqeq D$, where $C,D\in Concepts$.

In conclusion, concrete concepts are novel to $\sroiqc$ w.r.t. $\sroiq$, and it is through their use that the logic allows constraint modelling.

\paragraph*{A-Box}
The $\sroiq$ A-Box consists of a finite set of \textbf{assertions} of the form 
\begin{align}
C(a), \:\: R(a,b), \:\: \neg R(a,b), \:\: a\dot\neq b
\end{align}
Here $a,b\in N_{aI}$, $i_1,i_2\in N_{cI}$, $r\in Rel$ and $Rel$ is the set of relations defined for the constraint system. Allen's relation are a common choice for $Rel$, which enables us to define temporal constraint.

We extend the $\sroiq$ A-Box to include assertions of the form $(i_1\:r\:i_2)$ and $g(a,i_1)$, where $i_1,i_2\in N_{cI}$. 
We impose the constraint that for every $i\in N_I$ there must be some $g,a$ such that $ABox$ contains $g(a,i)$.
We assume Unique Name Assumption(UNA) for constraint individuals ($N_{cI}$), but not for abstract individuals ($N_{aI}$).


\subsection{Semantics}

An \textit{interpretation} $\mi$ is a tuple $(\Delta_\mi, \cdot^\mi, M_\mi)$, where $\Delta_{\mi}$ is the \textit{abstract domain}, $\cdot^\mi$ is the \textit{interpretation function}, and $M_\mi \in \mm$, is a complete constraint network of the constraint system $\mc$.

The interpretation function maps
\begin{itemize}
\item Each concept name $C\in N_C$ (atomic concept) to a subset $C^\mi$ of $\Delta_\mi$
\item Each abstract role $R\in N_{aR}$ to a subset $R^\mi$ of $\Delta_\mi \times \Delta_\mi$
\item Each concrete role $g\in N_{cR}$ to a subset $g^\mi$ of $\Delta_\mi \times V_{M_\mi}$
\item Each named abstract individual $a\in N_{aI}$ to an element $a^\mi\in\Delta_\mi$
\item Each constraint individual $i\in N_{cI}$ to an element $i^\mi\in V_{M_\mi}$
\end{itemize}
This notion is extended to other concepts and roles, as defined in the following paragraphs.


\paragraph{R-Box Interpretations}

The interpretation is extended to inverses as follows :
$(R^-)^\mi = \{\la x,y\ra \:\mid\: \la y,x\ra\in R^\mi\}$.
For a path $U=R_1\dots R_kg$ and $d\in \Delta_\mi$, $\:U^\mi(d)$ is defined as
\begin{align*}
\{x\in V_{M_\mi}\mid \exists  e_1,\dots,e_{k+1}:d=e_1,\la e_i,e_{i+1}\ra\in R_i^\mi \\ 
\text{ for } 1\leq i \leq k, \text{ and } \la e_{k+1}, x\ra\in g^\mi\}
\end{align*}

$\mi$ satisfies (is a model of) the concrete RIA $g_1\sqeq g_2$ iff $g_1^\mi\subseteq g_2^\mi$.
An interpretation $\mi$ satisfies the abstract RIA $R_1R_2...R_n \subseteq R$, if $R_1^\mi\circ R_1^\mi \circ ... \circ R_n \subseteq R^\mi$, where $\circ$ is teh binary composition operator.
The satisfaction of the role assertions is along the same lines as defined in \cite{sroiq}.

An interpretation is a model of R-Box if it satisfies all concrete and abstract RIAs and the role assertions.

\paragraph{Concepts}

The interpretation function is extended to arbitrary concepts as follows:
\begin{align*}
(\neg C)^\mi &:=  \Delta^\mi\setminus C^\mi,\\
o^\mi &:= \{a^\mi\},\text{ if } o=\{a\},\text{ else }\{x\}\text{ for }c\in\Delta_\mi\\
\top^\mi & := \dmi,  \\
\bot^\mi& :=  \emptyset, \\
(C\sqcap D)^\mi &:=  C^\mi \cap D^\mi,\\
(C\sqcup D)^\mi &:=  C^\mi \cup D^\mi,\\
(\exists R.C)^\mi &:=  \{d\in \Delta^\mi\mid \text{there is some }  d,e\in \Delta^\mi \\
&\text{ with }(d,e)\in R^\mi\text{ and }e\in C^\mi\},\\
(\forall R.C)^\mi &:=  \{d\in \Delta^\mi\mid \text{for all } e\in \Delta^\mi \text{ if }\\
&(d,e)\in R^\mi\text{ then }e\in C^\mi\},\\
(\exists R.Self)^\mi& :=  \{x\mid \la x,x \ra \in R^\mi\}, \\
(\geq nR.C)^\mi & :=  \{x\mid \#\{y.\la x,y \ra\in R^\mi\text{ and } y\in C^\mi\}\geq n \}, \\
(\leq nR.C)^\mi & :=  \{x\mid \#\{y.\la x,y \ra\in R^\mi\text{ and } y\in C^\mi\}\leq n \}  
\end{align*}
\begin{align*}
(\exists U_1,U_2.r)^\mi &:=  \{d\in \Delta^\mi\mid \text{there exist some } x_1\in U^\mi_1(d) \\  &  \text{ and }  x_2\in U^\mi_2(d) \text{ such that }  (x_1\:r\:x_2)\in M_\mi  \},\\
(\forall U_1,U_2.r)^\mi &:=  \{d\in \Delta^\mi\mid \text{for all }  x_1\in U^\mi_1(d)  \text{ and } \\ &  x_2\in U^\mi_2(d),  \text{ we have } (x_1\:r\:x_2)\in M_\mi \},\\
(\exists_c U_1,\{i\}.r)^\mi &:=  \{d\in \Delta^\mi\mid \text{there exist some }    x_1\in U^\mi_1(d) \\ &  \text{ and }  i\in N_{cI} \text{, such that } (x_1\:r\:i)\in M_\mi  \},\\
(\exists_c \{i\},U_2.r)^\mi &:=  \{d\in \Delta^\mi\mid \text{there exist some }   i\in N_{cI}  \\ & \text{ and } x_2\in U^\mi_2(d)  \text{, such that } (i\:r\: x_2)\in M_\mi  \},\\
(\forall_c U_1,\{i\}.r)^\mi &:=  \{d\in \Delta^\mi\mid \text{for all }  x_1\in U^\mi_1(d)  \text{ and for } \\ &  i\in N_{cI} \text{, we have } (x_1\:r\:i)\in M_\mi  \},\\
(\forall_c \{i\},U_2.r)^\mi &:=  \{d\in \Delta^\mi\mid  \text{for }   i\in N_{cI}  \text{ and all } \\ &  x_1\in U^\mi_2(d)  \text{, we have } (i\:r\: x_1)\in M_\mi  \},\\
(\leq_c n g)^\mi  &:=  \{x\mid \#\{c\mid\la x,c \ra\in g^\mi \text{ and } \\ & c\in V_{M_\mi}\}\leq n \}, \\
(\geq_c n g)^\mi & := \{x\mid \#\{c\mid\la x,c \ra\in g^\mi \text{ and } \\ & c\in V_{M_\mi}\}\geq n \}
\end{align*}

An interpretation $\mi$ is a \textit{model} of a concept $C$ iff $C^\mi\neq \emptyset$. 

\paragraph{T-Box interpretations}
$\mi$ is a model of a T-Box $\ct$ iff it satisfies $C^\mi\subseteq D^\mi$ for all $C \sqeq D$ in $\ct$.

\paragraph{A-Box interpretations}
An interpretation $\mi$ \textit{satisfies} (is a model of) an \textit{Abox} $\ma (\mi\models \ma)$ if for all individual assertions $\phi \in \ma$ we have $\mi \models \phi$, where
\begin{align*}
&\mi \models C(a) \: & if \: & a^\mi \in C^\mi; \\
&\mi \models a\dot\neq b \: & if \: &  a^\mi \neq b^\mi; \\
&\mi \models R(a,b) \: & if \: & \la a^\mi, b^\mi \ra \in R^\mi; \\
&\mi \models \neg R(a,b) \: & if \: & \la a^\mi, b^\mi \ra \notin R^\mi; \\
&\mi \models g(a,i) \: & if \: & a^\mi\in\dmi, i^\mi\in V_{M_\mi} \text{ and } \la a^\mi,i^\mi \ra\in g^\mi \\
&\mi \models (i_1 \: r \: i_2) \: & if \: & i_1^\mi,i_2^\mi \in V_{M_\mi}, r\in Rel; (i_1^\mi\:r\:i_2^\mi)\in M_\mi
\end{align*} 
The constraint individuals provide a way to directly address the variables of the constraint network, which would be formed by the tableau algorithm. It allows us to explicitly name concrete nodes of the completion system.

A knowledge base $\la \mt,\mr,\ma \ra$ is $satisfiable(consistent)$ if there exists an interpretation which is a model for each of $\mt,\mr$ and $\ma$.


\subsection{Inference Problems}

Consider the common inference problems.
\begin{enumerate}
\item KB Satisfiability : A knowledge base $\mk=\la \mt,\mr,\ma \ra$ is $satisfiable(consistent)$ if there exists an interpretation which is a model for each of $\mt,\mr$ and $\ma$.
\item Concept Satisfiability : A concept $C$ is satisfiable w.r.t. a knowledge base $\mk$ if there exists an interpretation which is a model of the KB and for which $C^{\mi}$ is not empty.
\item Concept Subsumption : A concept $C$ subsumed by a concept $D$ ($C \sqeq D$) if $C^{\mi} \subseteq D^{\mi}$ holds for all models of the $\mk$.
\item Instance Checking : An instance $a$ belonging to a concept $C$ is denoted as $C(a)$. The check succeeds if $a^\mi \in C^\mi$ for all models of $\mk$.
\end{enumerate}

Let $\mk$ be a $\sroiqc$ knowledge base and let $C,D$ be concepts in $\mk$. 
Let $a$ be an individual, such that $a\notin N_{aI}$ i.e. it is a named individual, new to the KB. 
Common inference problem can be converted to KB satisfiability as described below.
\begin{itemize}
\item Concept satsifiability : A concept $C$ is satisfiable iff $\mk\cup C(a)$ is satisfiable, for a novel $a$, not already mentioned in $\mk$.
\item Concept subsumption : $\mk\models (C\sqeq D)$ iff $\{a\}\sqeq (C\cap\neg D)$ is unsatisfiable.
\item Instance checking : the problem $\mk\models C(b)$ is equivalent to the concept subsumption problem $\{b\}\sqsubseteq C$, which is converted as mentioned above.
\end{itemize}

Thus a single tableau algorithm to decide the satisfiability of a $\sroiqc$ KB can be used to handle the above inference problems , as well.


\subsection{Knowledge Base Transformation}

The KB needs to be in a specific form before the Algorithm can operate. 
The following reductions are performed, in order, before the tableau algorithm operates on the KB.
The transformed, reduced KB is then input to the tableau algorithm.
We point out that all notions (e.g. $N_{aR}$ etc.) correspond to this reduced KB.

\paragraph{Elimination of Universal Role}
Let $C,D$ be concepts.
Let $U'\neq U$ be a role that does not occur in the knowledge base $\mk$. we modify the R-Box as defined below : 
\begin{align*}
&\mr^{U'}_{\:\:\:ah} := \mr_{\:ah} \cup \{R\sqeq U'\mid R \text{ occurs in }\mk\} \\
&\mr^{U'}_{\:\:\:a} := \mr_{\:a} \cup \{Tra(U'), Sym(U'), Ref(U')\}\\
&\mr^{U'} = \mr^{U'}_{\:\:\:ah} \cup \mr^{U'}_{\:\:\:a}
\end{align*}
Hereafter, when we mention $N_{aR}$, we do not include the universal role in it.

\paragraph{A-Box partial reduction}
We replace some of the $\ma$ assertions with an set of SGI axioms, to form a new T-Box $\ct_{A}$. Named individual, say $a$ used in assertions are converted to corresponding nominal set $o_a$. The assertions are converted to semantically equivalent axioms :
\begin{enumerate}
\item $C(a)$ to $o_a\sqeq C$
\item $R(a,b)$ to $o_a\sqeq \exists R.o_b$
\item $\neg R(a,b)$ to $\top\sqeq\forall R.\neg \{o_b\}(\{o_a\})$
\item $a\dot\neq b$ to $o_a \sqeq \neg o_b$
\end{enumerate}
The two new types of assertions introduced in $\gcsr$ are, however, not reduced.

\paragraph{$\calr_{\:a}$ transformation} This involves converting the role assertions into equivalent forms. The reduced RBox consists of only assertions of the form $Dis(S,S')$ and $Irr(R)$.

The role assertion $Ref(S)$ is converted to an equivalent T-Box axiom
\[\top \sqeq \exists S.Self\]
The role assertion $Fxnl(S),S\in N_{aR}$ is converted to an equivalent T-Box axiom
\[\top\sqeq\:\:\leq1S\]
The analogue for functional restriction on concrete role $g$ would be :
\[\top\sqeq\:\:\leq_c1g\]
The role assertions $Sym(R),Tra(R)$ is converted to an equivalent R-Box axiom
\begin{itemize}
\item $Sym(R)$: $R^- \sqeq R$
\item $Tra(R)$: $RR \sqeq R$
\end{itemize}

Next, we describe the translation of concepts into their NNF equivalents and the compilation of complex RIA into automata.

\subsubsection{Negation Normal Form (NNF)} 
The tableau algorithm expects the concept consructs to be in NNF.
The negation appears only in front of "primary" concepts in NNF. The set of \textit{primary concepts} consists of all atomic concepts ($N_C$ and $Nom$), $\exists S.Self$, $\exists_c U_1,U_2.r$, $\exists_c U_1,\{i\}.r$ (and its analogue $\exists_c \{i\},U_1.r$). For the other constructs, NNF equivalents are defined as below. In some of these cases, the NNF is not the logical equivalent, but preserves (un)satisfiability.

For the non-atomic primary concepts, NNF is not defined, for lack of a feasible closed expression. 
The semantics of these irreducible constructs are handled by the tableau algorithm by means of completion rules and special clash conditions, as mentioned later.
Some special constructs $(\exists_c U_1.q,\:\forall_c U_1.q)$ would be introduced by the tableau completion rules. NNF are not defined for these as they will never be present with a negation in front and are used for reasons internal to the tableau algorithm.

The NNF rules are shown :
\begin{align*}
\neg(\neg C) \:\rar&\: C\\
\neg(C\sqcap D) \:\rar&\: \neg C \sqcup \neg D \\
\neg(C\sqcup D) \:\rar&\: \neg C \sqcap \neg D \\
\neg(\exists R.C) \:\rar&\: \forall R.\neg C \\
\neg(\forall R.C) \:\rar&\: \exists R.\neg C \\
\neg(\leq n R.C) \:\rar&\: (\geq(n+1)R.C)\\
\neg(\geq (n+1) R.C) \:\rar&\: (\leq n R.C)\\
\neg(\geq 0 R.C) \:\rar&\: \bot \\
\neg(\leq n g) \:\rar&\: (\geq(n+1)g)\\
\neg(\geq_c (n+1) g) \:\rar&\: (\leq_c n g)\\
\neg(\geq_c 0 g) \:\rar&\: \bot \\
\neg(\forall_c U_1,U_2.r) \:\rar&\: \bigsqcup_{r' \in Rel,r'\neq r} \exists_c U_1,U_2.r'\\
\neg(\forall_c U_1,{i_1}.r) \:\rar&\: \bigsqcup_{r' \in Rel,r'\neq r} \exists_c U_1,{i_1}.r'\\
\end{align*}

\paragraph{RIA to automaton compilation}

This section is a direct adaptation of the work in \cite{sroiq} and is presented for completeness of the text. It describes the construction of a non-deterministic automaton for every complex (non-simple) role of the RBox, using the complex RIA's of $\mr_{ah}$.
Given a complex role $R$, the constructed automaton $\mb_R$ accepts all strings $\mu$ such that $\mu$ is a role chain which for all model interpretations $\mi$, satisfies : $\mu^\mi \subseteq R^\mi$. 
$\mb_R$ captures all implications between (paths of) roles and $R$ that are consequences of $\mr_{ah}$. 

The automata $\mb_{R}$ is built in steps.

Firstly, for each non-simple role name $R$ occurring in $\mk$, define an NFA $\ma_{R}$ as follows: $\ma_{R}$ contains a state $i_R$ and a state $f_R$ with the transition $i_R\stackrel{R}{\rightarrow}f_R$. The state $i_R$ is the only initial state and $f_R$ is the only final state. Moreover, for each $(w \sqeq R) \in \mr_{ah}$, $\ma_R$ contains the following states and transitions:
\begin{enumerate}
\item If $w=RR$, then $\ma_R$ contains $f_R\stackrel{\epsilon}{\rightarrow}i_R$, and
\item If $w=R_1\dots R_n$ and $R_1...R_n\neq R$, then $\ma_R$ contains
\[ i_R\stackrel{\epsilon}{\rightarrow}i_w\stackrel{R_1}{\rightarrow}f^1_w\stackrel{R_2}{\rightarrow}f^2_w\stackrel{R_3}{\rightarrow}\dots\stackrel{R_n}{\rightarrow} f^n_w \stackrel{\epsilon}{\rightarrow}f_R\]
\item If $w=RR_2\dots R_n$, then $\ma_R$ contains
\[ f_R\stackrel{\epsilon}{\rightarrow}i_w\stackrel{R_2}{\rightarrow}f^2_w\stackrel{R_3}{\rightarrow}f^3_w\stackrel{R_4}{\rightarrow}\dots\stackrel{R_n}{\rightarrow} f^n_w \stackrel{\epsilon}{\rightarrow}f_R\]
\item If $w=RR_2\dots R_n$, then $\ma_R$ contains
\[ f_R\stackrel{\epsilon}{\rightarrow}i_w\stackrel{R_2}{\rightarrow}f^2_w\stackrel{R_3}{\rightarrow}f^3_w\stackrel{R_4}{\rightarrow}\dots\stackrel{R_n}{\rightarrow} f^n_w \stackrel{\epsilon}{\rightarrow}f_R\]
\item If $w=R_1\dots R_{n-1}R$, then $\ma_R$ contains
\[ i_R\stackrel{\epsilon}{\rightarrow}i_w\stackrel{R_1}{\rightarrow}f^1_w\stackrel{R_2}{\rightarrow}f^2_w\stackrel{R_3}{\rightarrow}\dots\stackrel{R_{n-1}}{\rightarrow} f^{n-1}_w \stackrel{\epsilon}{\rightarrow}i_R\]
\end{enumerate}
where all $f^i_w,i_w$ are assumed to be distinct. 

To construct a mirror copy $\ma_{R^-}$ of an automaton $\ma_R$, the following procedure is followed:
\begin{itemize}
\item Make final states to non-final states
\item Make initial states to non-initial but final states
\item Replace each transition $p\stackrel{S}{\rightarrow}q$ with $q\xrightarrow{Inv(S)}p$, where $S$ is a (possibly inverse) role.
\item Replace each transition $p\stackrel{\epsilon}{\rightarrow}q$ with $q\xrightarrow{\epsilon}p$
\end{itemize}

Secondly, we define the NFAs $\hat{\ma_R}$ as follows:
\begin{itemize}
\item If $R^- \;\dot\sqeq \;R \;\notin \;\mr$ then $\hat{\ma_R}:=\mr_a$
\item If $R^- \;\dot\sqeq \;R \;\in \;\mr$ (R is transitive), then $\hat{\ma_R}$ is obtained as follows:
\begin{enumerate}
\item Take a disjoint union of $\ma_S$ with a mirrored copy of $\ma_S$
\item Make $i_R$ the only initial state and $f_R$ the only final state
\item For $f'_R$ the copy of $f_R$ and $i'_R$ the copy of $i_R$ the copy of $i_R$, add transitions $i_R\stackrel{\epsilon}{\rightarrow}f'_R$,
$f'_R\stackrel{\epsilon}{\rightarrow}i_R$ 
\end{enumerate} 
\end{itemize}

Thirdly, the NFAs $\mb_R$ are defined inductively over $\prec$:
\begin{itemize}
\item If $R$ is minimal w.r.t. $\prec$ (i.e., there is no $R'$ with $R'\:\prec\:R$), we set $\mb_R:=\hat{\ma_R}$
\item Otherwise, $\mb_R$ is the disjoint union of $\hat{\ma_R}$ with a copy $\mb'_S$ of $\mb_S$ for each transition $p\xrightarrow{S}q$ in $\hat{\ma_R}$ with $S \neq R$. Moreover, for each such transition, we add $\epsilon$-transistions from $p$ to the initial state in $\mb'_S$ and from the final state in $\mb'_S$ to $q$, and we make $i_R$ the only initial state and $f_R$ the only final state $\mb_R$
\end{itemize}

We will assume that final, reduced and converted KB formed by following the above series of operations if given by $\mk=\la\mr,\mt,\ma\ra$


\subsection{An Augmented Tableau for $\sroiqc$}\label{augtab}

We propose an \textbf{augmented tableau} for $\sroiqc$, $T_A = \la T, N \ra $ where 
$T=\la \vat,\vct,\ml^T,\eat,\ect\ra$  is a tableau and $N$ is a constraint network with $V_N = \vct$. 
The augmented tableau for a KB is a graphical representation of a model of the KB $\mi=(\Delta_\mi,\cdot^\mi, M_\mi)$. 

The elements of the non-empty set $\vat$ forms the nodes of tableau $T$. 
$Nom(T)\subset \vat$ is the set of nominal nodes and represents named individuals.
$\vct$ forms the variables which participate in constraint relations that compose $N$.
Further, we define the following mappings :
\begin{itemize}
\item $\ml^T: \vat \rar 2^{clos(\ct,\calr)}$
\item $\eat: N_R \rar 2^{\vat \times \vat}$
\item $\ect: N_{cF} \rar 2^{\vat \times \vct}$
\end{itemize}

\noindent
For a concept $C$ and $D$, and a $\sroiqc$ KB $\mk = \la \ma,\mt,\mr\ra$, we define $sub(D)$, the set of \textit{sub-expressions} of $D$. 
The intuitive significance of $sub(D)$ is in describing the contents of the label sets of graph nodes i.e. the concepts formed that label the nodes of graph, if the KB consists of no other concept other than $D$. Consider the following :
\begin{enumerate}
\item 
If $D$ is of the form $\neg C,\exists R.C, \forall R.C,\leq nR.C$ or $\geq nR.C$, then $sub(D)=\{C,D\}$
\item
If $D$ is of the form $C_1 \sqcap C_2$ or $C_1 \sqcup C_2$, then $sub(D) = \{D,C_1,C_2\}$
\item 
Otherwise $sub(D) = \{D\}$  
\end{enumerate}
$Sub(D)$ is the closure over the sub-expressions of $D$ and their negations (NNF forms) \cite{Horr97b}.
For a concept $C$, and $\mk=\la\mt,\mr,\ma\ra$, we define :
\begin{align*}
clos(C,\mr)&:= Sub(C) \cup \{\forall \mathcal{B}_S(q).D \mid \forall S.D \in Sub(C)\\
&\text{and }\mathcal{B}_S \text{ has a state q }\}\\
clos(\ct,\mr)&:= \bigsqcup_{C\sqeq D \in \ct} clos(NNF(\neg C \sqcup D),\mr)
\end{align*}

For the following we assume $s,t\in \vat; C,C_1,C_2\in clos(\ct,\mr)$, unless mentioned otherwise. Further, for a role $S\in N_R$, we define the set of successors of a node $s$ w.r.t. $S$, with $C$ in their label set as :
\[S^T(s,C) := \{t\in\vat \mid \la s,t\ra\in \eat(S); C \in \ml(t)\}, \]
For a path $U=R_1...R_kg$, a node $y\in\vct$ is called a \textit{U successor} of a node $x\in\vat$ if there exist $e_1,\dots,e_{k+1}\in\vat$ such that $\la x,e_1\ra\in\eat(R_1)$, $\la e_i,e_{i+1}\ra\in\eat(R_i)$ for $2\leq i\leq k-1$ and $\la e_k,y\ra\in\ect(g)$.

A valid augmented tableau $T_a=\la T,N\ra$, for $\mk$, must satisfy the propositions below.
Propositions relating to abstract domain constructs (denoted by $P_a$) are borrowed from $\sroiq$ and included here for completeness. The propositions relating to the concrete domain constructs are denoted by $P_c$. 
\begin{itemize}

\item
$(P_a\:\sqsubseteq)$: If $(C\sqsubseteq D)\in\mathcal{T}$, then $NNF(\neg C\sqcup D)\in\lt(s)$ for all $s\in\vat$

\item
$(P_a\:\neg)$: If $C \in \lst$, then $\neg C \notin \lst$ ($C$ atomic or $\exists R.Self$ or, $\top$ or, $\bot$)

\item \textit{(P$_a$\:Self)}: If $\exists S.Self\in \lst$, then $\la s,s \ra\in \eat(S)$ 

\item \textit{(P$_a\neg$\:Self)}: If $\neg\exists S.Self\in \lst$, then $\la s,s \ra\notin \eat(S)$  

\item \textit{(P$_a\:\sqcap$)}: If $C_1\sqcap C_2\in \lst$, then $C_1\in \lst$ and $C_2\in \lst$  

\item \textit{(P$_a\:\sqcup$)}: If $C_1\sqcup C_2\in \lst$, then $C_1\in \lst$ or $C_2\in \lst$

\item \textit{(P$_a\:\exists$)}: If $\exists R.C\in\lst$ then there is some $t$ with $\la s,t \ra \in \eat(R)$ and $C \in \ltt$

\item \textit{(P$_a\:\forall$)}: If $\forall R.C\in\lst \text{ and } \la s,t \ra \in \eat(R) \text{ then  } C \in \ltt$

\item \textit{(P$_a$\:Inv)}: $\la x,y\ra \in \eat(R)$ iff $\la y,x\ra \in \eat(Inv(R))$   

\item \textit{(P$_a \:\leq$)}: If $(\leq n S.C)\in \lst$, then $\#S^T(s,C) \leq n$    

\item \textit{(P$_a \:\geq$)}: If $(\geq n S.C)\in \lst$, then $\#S^T(s,C) \geq n$    

\item \textit{(P$_a\:\leq$ Neighbour)}: If $(\leq n S.C)\in \lst$, and $\la s,t \ra\in \eat(S)$, then $C\in \ltt$ or $nnf(\neg C)\in \ltt$ 

\item \textit{(P$_a$\:Dis)}: If $Dis(R,S)\in \mr_a$, then $\eat(R)\cap \eat(S) = \emptyset$

\item \textit{(P$_a$\:Ref)}: If $Ref(S)\in \mr_a$, then $\la s,s \ra \in \eat(S)$ for all $s\in \vat$ 

\item \textit{(P$_a$\:Subrole)}: If $\la s,t \ra \in \eat(R)$ and $R \sqeqs S$ then  $\la s,t \ra \in \eat(S)$

\item \textit{(P$_a$\:Nom)}: $o \in \lst$  for some $s \in \vat$, for each $o \in Nom \cap clos(C_0,\mr)$

\item \textit{(P$_a$\:NomUnq)}: If $o \in \lst\cap\ltt$ for some $o \in Nom$ then $s=t$

\item \textit{(P$_a$\:$\mb\:transition$)}: if $\forall \mb(p).C\in\ml(s)$, $\la s,t \ra\in\eat(R)$, and $(p\overset{R}{\longrightarrow}q)\:\in\mb(p)$, then $\forall\mb(q).C\in\lt(t)$

\item \textit{(P$_a$\:$\mb\:\epsilon$)}: if $\forall \mb.C\in\ml(s)$, and $\epsilon\in L(\mb)$, then $C\in \lt(s)$

\item \textit{(P$_c\:\forall_cU$)}: If $(\forall_c U_1,U_2.r)\in \lst$, then for all $x,y$ such that $x$ is a $U_1$ successor and $y$ is a $U_2$ successor of $s$, we have $(x\:r\:y)\in N$.

\item \textit{(P$_c\:\exists_cU$)}: If $(\exists_c U_1,U_2.r)\in \lst$, then there must exist $x,y\in\vct$ such that $x$ is a $U_1$ successor and $y$ is a $U_2$ successor of $s$ and $(x\:r\:y)\in N$.

\item \textit{(P$_c\:\forall_ci$)}: If $(\forall_c U_1,\{i\}.r)\in \lst$, then for all $x$ such that $x$ is a $U_1$ successor of $s$ and $i\in N_{cI}$, we have $(x\:r\:i)\in N$. The symmetric case of $(\forall_c\{i\},U_1.r)$ is handled analogously.

\item \textit{(P$_c\:\exists_ci$)}: If $(\exists_c U_1,\{i\}.r)\in \lst$, then there must exist $x\in\vct$ such that $x$ is a $U_1$ successor $s$ and $(x\:r\:i)\in N$, where $i\in N_{cI}$. The symmetric case of $(\exists_c\{i\},U_1.r)$ is handled analogously.

\item \textit{(P$_c\:\neg\exists_cU$)}: If $(\neg\exists_c U_1,U_2.r)\in \lst$, then there must not exist any $y,y'\in\vct$ such that $y$ is a $U_1$ successor of $s$, $y'$ is a $U_2$ successor of $s$ and $(y\:r\:y')\in N$.

\item \textit{(P$_c\:\neg\exists_ci$)}: If $(\neg\exists_c U_1,\{i\}.r)\in \lst$, then there must not exist any $y\in\vct$ such that $y$ is a $U_1$ successor of $s$, and $(y\:r\:i)\in N$.

\item \textit{(P$_c \:\leq_c$)}: If $(\leq n g)\in \lst$, then 
$\#\{c_i \mid c_i\in \vct, \la s, c_i\ra\in\ect(g)\}\leq n$.

\item \textit{(P$_c \:\geq_c$)}: If $(\geq n g)\in \lst$, then  
$\#\{c_i \mid c_i\in \vct, \la s, c_i\ra\in\ect(g)\}\geq n$.

\item \textit{(P$_a\:Subrole_c$)}: If $\la y_1,y_2 \ra \in \eat(g)$ and $g \sqeqs g'$ then  $\la y_1,y_2 \ra \in \eat(g')$

\item \textit{(P$_c\:N$Sat)}: $N$ should be satisfiable.

\end{itemize}

\paragraph{Proposition :}
\textit{There exists an augmented tableau for a $\sroiqc{}$ KB iff there exists a model for it.}


\section{Tableau Algorithm for SROIQ(C)}

The Tableau algorithm generates a \textbf{completion system} $\mathcal{S} = ( \mg,\mn,\cts)$, where \\
$\mg =(\vas,\vcs,\eas,\ecs,\ls,\mss,\dot\neq)$ is a \textbf{completion graph}, $\cn$ is a finite \textbf{constraint network} with $V_{\cn}=\vcs$ and $\mathcal{Q}$ is the \textbf{constraint template set}.
$\vas$ is the set of abstract nodes, $\vcs$is the set of concrete nodes.
$Nom(\ms)\subset \vas$ is the set of nominal nodes and represent named individuals.
$\mss$ labels each of the concrete nodes of $\cs$ with a set of marker symbols.
$\eas$ is set of abstract edges of the form $(a,b)$, where $a,b\in\vas$.
$\ecs$ is set of concrete edges of the form $(a,c)$, where $a\in\vas,c\in\vcs$.

The completion system is a finite graphical representation of a (possibly infinite) tableau, with $\ls$ labelling nodes with concepts, and edges with roles.


\subsection{Preliminaries}
Here we discuss the terminology required to introduce the tableau algorithm.
Let KB be a $\sroiqc$ knowledge base consisting of $\la \ma,\mr,\ct \ra$. 
The algorithm assumes that the KB has been reduced.
The algorithm is for deciding the satisfiability of a $\sroiqc$ KB. 
However, other common inference problems can be reduced to KB satisfiability, as described earlier.

\paragraph{Neighbours and Successors}

If $\la x, y\ra \in \eas\:or\:\ecs$, then $y$ is called a \textit{successor} of $x$, and $x$ is called a \textit{predecessor} of $y$. Further, if $R \in\ls(\la x,y \ra)$, then $y$ is the successor of $x$ wrt $R$. \textit{Ancestor} is the transitive closure of predecessor, and \textit{descendant} is the transitive closure of successor.

For $x,y\in\vas,R\in N_{aR}$, $y$ is called an $R$-\textit{successor} of a node $x$ if, for some $R'\in N_{aR}$ with $R' \sqeqs R,\: R'\in\ls(x,y)$.
Similarly, for $x\in\vas,y\in\vcs,g\in N_{cR}$, $y$ is called an $g\:successor$ of a node $x$ if, for some $g'\in N_{cR}$ with $g' \sqeqs g, g'\in\ls(x,y)$. 
It may be noted that this notion is different from that of successor wrt R, defined above.

A node $y$ is called a \textit{R-neighbour} of a node $x$, if $y$ is a $R$-successor of $x$ or if $x$ is a $Inv(R)$-successor of $y$.
That is, $x$ has an outgoing $R$ edge going towards $y$, or has a incoming $Inv(R)$ edge coming from $y$.

For a path $U=R_1\dots R_kg$, a node $c\in\vcs$ is called a \textit{U-successor} of a node $a\in\vas$ if there exist $e_1,\dots,e_{k}\in\vas$ such that $e_1$ is the successor of $a$ w.r.t $R_1$, $e_{i}$ is the successor of $e_{i-1}$ w.r.t $R_i$ for $2\leq i\leq k$ and $c$ is the $g$ successor of $e_k$.
For a path $U=R_1\dots R_kg$, a node $c\in\vcs$ is called a \textit{U-successor} of a node $a\in\vas$ if there exist $e_1,\dots,e_{k}\in\vas$ such that $e_1$ is the $R_1$ neighbour of $a$ , $e_{i}$ is the $R_i$ neighbour of $e_{i-1}$ for $2\leq i\leq k$ and $c$ is the $g$ successor of $e_k$.
Unlike in completion system, we do not maintain any distinction between successor and neighbour in the augmented tableau because, as elaborated later, in the tableau all subrole and inverse related information is made explicit.

\paragraph{Marker Symbols and Constraint Template Set}

The \textit{constraint template set} $\cts$ is a set of expressions of the form $(q_{1}\:r\:q_{2})$ \textit{(the positive template)} or $\neg(q_{1}\:r\:q_{2})$ \textit{(the negative template)}, where $q_{1},q_{2}$ are marker symbols and $r\in Rel$.
It may be noted that it is these marker symbols which label a concrete node using $\mss$.
Let $c_1,c_2$ be any two concrete nodes with $q_1\in\mss(c_1),q_2\in\mss(c_2)$. 
If $(q_1\:r\:q_2)\:\in\:\cts$, then the completion rules add a constraint $(c_1\:r\:c_2)$ to $\mn$. 
In comparison, if $\neg(q_1\:r\:q_2)\in\cts$, then the presence of $(c_1\:r\:c_2)\in\mn$ leads to a clash.
The negative template is used with clash conditions and completion rules to ensure the semantics of concepts of the form $(\neg\exists U_1,U_2.r)$ or $(\neg\exists U_1,\{i\}.r)$ or $(\neg\exists \{i\},U_1.r)$.

\paragraph{Blocking}

Without a blocking mechanism, the completion rules may produce a infinite completion graph composed of repeating units, which would compromise the termination of the tableau algorithm.
The presented blocking scheme is an extension of the one used for $\sroiq$.

A non-nominal abstract node of $\ms$ is called a $blockable$ node.
In order for blocking to occur, a series of blocking checks need to be passed successfully.
We take up them in the following paragraphs.

\paragraph{Blocking Check 1 (BC-1)}
Let $a,b,a_p,b_p\in\vas$.
$a$ and its descendant $b$ are said to pass BC-1, if all of the following conditions are met :
\begin{enumerate}
\item $a_p$ is a predecessor of $a$, $b_p$ is a predecessor of $b$ and $a,a_p,b_p$ are the ancestors of $b$
\item $a,a_p,b,b_p$ are distinct, and all nodes on the path from $a$ to $b$ are blockable
\item $\ls(a)=\ls(b)$ and $\ls(a_p)=\ls(b_p)$
\item $\ls(\la a_p,a \ra)=\ls(\la b_p,b \ra)$
\end{enumerate}
This check ensures that there is indeed a repetition in the abstract node backbone.
This check is the direct adaptation from $\sroiq$ and is required since $\sroiqc$ contains all the constructs of $\sroiq$.
The distinctness of the nodes just ensures that there is at least one node (and two edges) between the blocker and the directly blocked node, and is used for arguing the correctness of the algorithm intuitively.

\paragraph{Blocking Check 2 (BC-2)}

For the above mentioned nodes, we define the following :
\begin{align*}
relevantCNodes(a)&=\{c \:\mid\: c\in\vcs;\: \text{$c$ is the successor of $a$ or $a_p$ } \}\\
relevantCNodes(b)&=\{c \:\mid\: c\in\vcs;\: \text{$c$ is the successor of $b$ or $b_p$ } \}\\
relevantCNet(a)&=\{(c\:r\:c') \:\mid\: \{c,c'\}\subseteq relevantCNodes(a);\: (c\:r\:c')\in\mn \}\\
relevantCNet(b)&=\{(c\:r\:c') \:\mid\: \{c,c'\}\subseteq relevantCNodes(b);\: (c\:r\:c')\in\mn \}\\
\end{align*}
We define an injective mapping $\phi\::\:relevantCNodes(a)\rar relevantCNodes(b)$ such that 
\begin{itemize}
\item
for every $c\in relevantCNodes(a)$ which is the successor of $a_p$, $\ls(\la a_p,c \ra)=\ls(\la b_p,\phi(c) \ra)$
\item
for every $c\in relevantCNodes(a)$ which is the successor of $a$, $\ls(\la a,c \ra)=\ls(\la b,\phi(c) \ra)$
\item
for every $c\in relevantCNodes(b)$ which is the successor of $b_p$, $\ls(\la b_p,c \ra)=\ls(\la a_p,\phi^-(c) \ra)$
\item
for every $c\in relevantCNodes(b)$ which is the successor of $b$, $\ls(\la b,c \ra)=\ls(\la a,\phi^-(c) \ra)$
\end{itemize}
In the above discussion $\phi^-$ is the inverse mapping of $\phi$.

A blockable node $a$ and its descendant $b$ pass BC-2 if :
\begin{itemize}
\item
they pass BC-1
\item
the mapping $\phi$ exists
\item
for every $c,c'\in relevantCNodes(a)$ : $(c \:r\: c')$ iff $(\phi(c) \:r\: \phi(c'))$
\item
for every $c,c'\in relevantCNodes(b)$ : $(c \:r\: c')$ iff $(\phi^-(c) \:r\: \phi^-(c'))$
\end{itemize}

This check is required to ensure that when the blocked region is unravelled into an infinite structure while constructing the tableau, then the semantics of the concrete concepts are indeed enforced.
We return to this in the soundness proof.

\paragraph{Blocking Check 3 (BC-3)}

A node $a_2$ is a $strict$ $descendant$ of $a_1$, if it is a non-nominal descendant of $a_1$ and can be reached from $a_1$ without encountering a nominal node on the connecting path.
This notion of "strict descendant" differs from "descendant" because unlike the latter, it does not transitively extend to nodes whose path to $a_1$ has a nominal in it.

For nodes $a$ and $b$, which pass BC-2, we define the following terms :
\begin{align*}
inter&nalCNodes(a,b)\:=\{c\:\mid\:c\in\vcs;\:c\text{ is a strict}\\
&\text{descendant of } a \text{ but not of } b\}\\
exter&nalCNodes(a,b) = \{ c \:\mid\: c\notin internalCNodes(a,b); \\
&c\notin strict\_descendant(b); \text{ for some } c'\in internalCNodes(a,b) \\
&\text{ either } \la c,r,c' \ra\in\mn \text{ or }\la c',r,c \ra\in\mn \}\\
assoc&iatedCNodes(b)=\{c \:\mid\: \text{$c\in\vcs$ is the successor of $b_p$ } \}\\
assoc&iatedCNodes(a)=\{c \:\mid\: \text{$c\in\vcs$ is the successor of $a_p$ } \}\\
relat&iveCNodes(a,b) = associatedCNodes(a)\\
fixed&CNodes(a,b) = externalCNodes(a,b) - relativeCNodes(a,b)\\
\end{align*}

We could alternatively define $fixedCNodes(a,b)$ as :
\begin{align*}
fixed&CNodes(a,b) = \{c \:\mid\: c\text{ is the successor of a nominal node $r$; and} \\
&\text{for some $c'\in internalCNodes(a,b)$ either } \la c,r,c' \ra\in\mn \text{ or }\la c',r,c \ra\in\mn \}
\end{align*}

Though the requirement for the above definitions will become clearer in the soundness proof, we explain the intuition behind the above sets.
If $a$ and $b$ pass the remaining tests, and $a$ blocks $b$, then the nodes between them would form a "unit".
Units isomorphic / similar in structure to this unit would get repeated over and again to form an infinite structure when the completion system is unravelled into a tableau.
The concrete nodes belonging to one such unit in the tableau would map back to the nodes of $internalCNodes$.
The concrete nodes of one such unit may enter into constraints with the nodes of the unit immediately before it.
These nodes of the preceding unit map back to $associatedCNodes(b)$.
Also, sometimes, the nodes of different units all form constraints with the same node, external to all of them.
The set of such external concrete nodes maps back to $fixedCNodes(a,b)$.

We define :
\begin{align*}
cNetNodes&(a)= associatedCNodes(a)\cup fixedCNodes(a,b)\\
cNetNodes&(b)= associatedCNodes(b)\cup fixedCNodes(a,b)
\end{align*}

We define a mapping $\theta:cNetNodes(a)\rar cNetNodes(b)$, as follows :
\begin{equation*}
\theta(c) \: =
\begin{cases}
\phi(c)& \text{if } c\in associatedCNodes(a)
\\
c& \text{if } c\in fixedCNodes(a,b)
\end{cases}
\end{equation*}
On account of BC-2, this exists.

We call the constraint networks formed by variables in $cNetNodes(a)$ as $cNet(a)$ and those by the variables in $cNetNodes(b)$ as $cNet(b)$.
$cNet(a)$ and $cNet(b)$ are said to be \textbf{isomorphic} if for $c,c'\in cNetNodes(a)$ and $(c,r,c') \in \mn$, we have $(\theta(c),r,\theta(c')) \in\mn$.

$ $\\
A blockable node $a$ and its descendant $b$ pass BC-3 if the following holds :
\begin{itemize}
\item
they pass BC-2
\item
the constraint networks $cNet(a)$ and $cNet(b)$ are complete and isomorphic.
\end{itemize}

If $a$ and its descendant $b$ pass all the checks, $a$ is said to \textbf{directly block} $b$.
$a$ is termed the blocker, $b$ and its descendants are said to be \textbf{blocked}.
The notion of indirectly blocked is as defined in $\sroiq$.

This check is required to be able to apply the patchwork property of the constraint system, in order to argue for the satisfiability of the constraint network of the unravelled augmented tableau.


\paragraph {Merging and pruning}

The merging carried out here is similar to \cite{sroiq}. \\

The following steps must be carried out when merging abstract node $y$ into $x$ :
\begin{enumerate}
\item If $\la x,c\ra\in\ecs$, add $g$ to $\ls(\la x,c\ra)$ i.e. if $x$ already has a edge with $c$ (only way this could happen is if $c\in N_{cI}$), then the label set of this edge is augmented with that from $\la y,c \ra$.
\item Else, create a new edge between $x$ and $c$, and set $\ls(\la x,c \ra)=\ls(\la y,c \ra)$
\end{enumerate}
Informally, we move all the concrete node successors of $y$ to now be the successors of $x$.

We point out that while merging an abstract node ($a$) into another ($b$), we drop/remove all the abstract edges outgoing from $a$; but we have, however, retained the concrete edges.
This is required to avoid a threat to termination.
Consider that a abstract successor to node $a$ w.r.t $R$, $a_s$, created by completion rules acting on $(\exists_c Rg,g.r)\in\ls(a')$.
Assume $c$ is the $g$ successor of $a_s$.
If it so happens that $a_s$ is merged back into $a$, then if we drop the concrete edges then we return to exactly the same state that existed before the completion rule acted on $(\exists_c Rg,g.r)\in\ls(a')$, and this "loop" could compromise termination.
Hence, we retain the concrete edges and nodes.

Consider the case of merging a concrete node $c_1$ into another concrete node $c_2$, where there exists no relation $c_1\dot\neq c_2$.
The constraints of the representation scheme ensure that $c_1$ and $c_2$ are the successors of the same abstract node, and are not both constraint individuals simultaneously.
The following steps must be performed :
\begin{enumerate}
\item 
For $a\in\vas$ such that $(a,c_1)\in\ecs$, the following steps are performed :
\begin{enumerate}
\item Set $\ls(\la a,c_2\ra) = \ls(\la a,c_2\ra)\cup\ls(\la a,c_1\ra)$
\item Remove $\la a,c_1 \ra$ from $\ecs$
\end{enumerate}
\item 
Set $\mss(c_2) = \mss(c_1) \cup \mss(c_2)$
\item 
Rename $c_1$ to $c_2$ in all constraints in $\mn$.
\item 
Add $c_2 \dot\neq c'$ for all $c'$ such that $c' \dot\neq c_1$
\end{enumerate}

Pruning is done similar to $\sroiq$, but when an abstract node is pruned, the concrete nodes ($\notin N_{cI}$) successors of the nodes are also pruned from $\ms$.


\paragraph{Clash Conditions}

For completeness, the clash conditions of $\sroiq$ are mentioned here. These are adapted directly into our algorithm, since $\sroiqc$ contains all constructs of $\sroiq$.
\begin{enumerate}
\item There exists $a\in\vas$, such that $\bot\in\ls(a)$
\item There exists $a\in\vas$, such that $\{C,\neg C\}\subseteq \ls(a)$
\item $\neg \exists S.Self\in\ls(a)$ and $S\in\ls(\la a,a \ra)$.
\item There is some $Dis(S,S')\in R_a$ and for some $a,b\in\vat$, $b$ is both $S$ and $S'$ neighbour of $a$.
\item There is some concept $\leq nS.C\in\ls(a)$, and there exist more than $n$ $S\:\:$neighbours of $a$.
\item For some nominal $o\in Nom$, $o\in\ls(a)\cap\ls(b)$ for distinct $a\dot\neq b$.
\end{enumerate}

Apart from the above, the completion system is said to contain a $clash$ in either of the following cases :
\begin{enumerate}
\item There exists $a\in\vas$, such that $\leq_c ng\in\ls(a)$, and there exist $c_1\dots c_k\in\vcs$ such that $k>n$, $c_i\dot\neq c_j$ for $1\leq i < j\leq k$ and each $c_i$ is a $g$ successors of $a$ for $1\leq i < k$
\item There exist $c_1,c_2\in\vcs$, such that $(c_1\:r\:c_2)\in\mn$ even though $\neg(q_1\:r\:q_2)\in\cts$ for some $q_1\in\mss(c_1)$, $q_2\in\mss(c_2)$.
\item $\cn$ is not satisfiable
\end{enumerate}


\subsection{Algorithm Initialization}

If $o_1,\dots,o_K\in Nom$, then the tableau algorithm starts with the completion system $\mathcal{S} = ( \mg,\mn,\cts)$, where $\mg =(\vas,\vcs,\eas,\ecs,\ls,\mss,\dot\neq)$. 
Here, $\vas$ = $\{r_1,\dots,r_K\}$, where $\ls(r_k)=\{o_k\}$ for $1\geq k \geq K$, and $o_k=\{p_k\}$ for $p_k\in N_{aI}$ being an named individual.
In case there are no nominals in the KB, then a seed node $r_0$ is the sole member of $\vas$.

For every A-Box assertion of the form $g(p_k,i_l)$, where $p_k\in N_{aI},i_l\in N_{cI},\{p_k\}=o_k$ and $o_k\in\ls(r_k)$, do the following : add a concrete node named $i_l$ to $\vcs$, add an edge $\la r_k,i_l \ra$ to $\ecs$ and set $\ls(\la r_k,i_l \ra)=\{g\}$.
Further, for every assertion of the form $(i_1\:r\:i_2)$, add $(i_1\:r\:i_2)$ to $\mn$.
For all such concrete nodes $i_1\dots i_L\in N_{cF}\cap \vcs$, add $i_l\dot\neq i_m$ for $1\leq (l, m)\leq k$.

$\cg$ is then expanded by repeatedly (and non-deterministically) applying the applicable expansion(completion) rules. 
This stops if either a $clash$ occurs or if no more rules are applicable, in which case $\cs$ is said to be $complete$.
The completion system $\cs$, if complete and clash-free, can be unravelled to form an augmented tableau for the Knowledge Base $\la\ma,\ct,\calr \ra$.
The algorithm returns a complete and clash-free completion system, iff the KB is consistent.

\subsection{The Completion Rules}

For completeness we first mention the completion rules of $\sroiq$, all of which are valid completion rules in the tableau procedure for $\sroiqc$.

Unless mentioned otherwise, assume, $C,D,C_1,C_2$ are concepts. $a,b\in\vas$; $c,c_1,c_2\in\vcs$; $i,i_1,i_2\in N_{cI}$; $R,R'\in N_{aR}$ are abstract roles; $S,S'$ are simple abstract roles; $U_1,U_2$ are paths; $g,g'\in N_{cR}$ are concrete roles. $G,G'$ are paths, role chains or $\epsilon$. $q$ is a $marker$ and $q_s$ is a marker symbol new to $\ms$.
When creating a new node $a$ which is an $R$ successor of existing node $b$, we use $\ls(\la b,a\ra)=\{R\}$ to imply $\vas=\vas\cup\{a\}$ and $\eas=\eas\cup\{\la b,a\ra\}$ in addition to $\ls(\la b,a\ra)=\{R\}$. Analogously, for concrete node creation.

\begin{itemize}
 
\item
\textbf{R$\sqcap$}: if $C_1\sqcap C_2 \in \ls(a),\:a$ is not blocked, and $\{C_1,C_2\}\nsubseteq \ls(a)$, \\
Then $\ls(a) \rar \ls(a) \cup \{C_1,C_2\}$

\item
\textbf{R$\sqcup$}: If $C_1\sqcup C_2 \in \ls(a),\:a$ is not blocked, and $\{C_1,C_2\}\cap\ls(a)=\emptyset$, \\
Then $\ls(a) \rar \ls(a) \cup \{E\}$ for some $E\in \{C_1,C_2\}$

\item
\textbf{R$\:gci$}: If $C\sqsubseteq D\in\ct$ and $(\neg C\cup D) \notin \ls(a)$,\\
Then set $\ls(a):=\ls(a)\cup NNF(\neg C\cup D)$

\item
\textbf{R$\exists$}: If $\exists S.C \in \ls(a)\:,a$ is not blocked, and $a$ has no $R$-neighbour $b$ with $C\in\ls(b)$\\
Then create a new node $b$ with $\ml(\la a,b \ra) := \{R\} \text{ and } \ls(b) := \{C\}$

\item
\textbf{R Self-Ref}: If $\exists S.Self \in \ls(a)\text{ or }Ref(S)\in \mr_a,\:a$ is not blocked, and  $S\notin \ml(\la a,a \ra)$\\
Then add an edge $\la a,a \ra$ if it does not exist yet and set $\ls(\la a,a \ra)\rar \ls(\la a,a \ra)\cup \{S\}$

\item
\textbf{R$\forall_1$}: if $\forall S.C \in \ls(a),\:a$ is not indirectly blocked, and $\forall \mb_S.C\notin\ls(a)$\\
then $\ls(a) \rar \ls(a) \cup \{\forall \mb_S.C \}$ 

\item
\textbf{R$\forall_2$}: If $\forall \mb(p).C \in \ls(a),\:a$ is not indirectly blocked, and 
 $p\stackrel{S}{\rightarrow}q\text{ in }\mb(p)$ 
  and there is an $S$-forward-neighbor $b$ of $a$ with $\forall \mb(q).C \notin \ls(b)$\\
then $\ls(b) \rar \ls(b) \cup \{\forall \mb_S(q).C \}$ 

\item
\textbf{R$\forall_3$}: If \: $\forall \mb.C \in \ls(a),\:a$ is not indirectly blocked, 
 $\epsilon \in L(\mb), \text{ and } C \notin \ls(a)$ \\
then $\ls(a) \rar \ls(a) \cup \{C\}$ 

\item
\textbf{R choose-rule}: If \: $(\leq nS.C)\in \ls(a),\:a$ is not blocked, and 
 there is an $S$-neighbor $y$ of $x$ with $\{C,\dot\neg C\}\cap \ly=\emptyset$\\
 then $\ly\rar\ly\cup\{E\} \text{ for some } E \in \{C,\dot\neg C\}$

\item
$\geq$-rule: If 1. $(\geq n S.C)\in \ls(a), x$ is not blocked, and\\
2. There are not $n$ safe $S$-forward-neighbors $y_1,\dots,y_n$ of $a$ with 
 $C \in \ml(y_i)\text{ and } y_i\dot\neq y_j\text{ for }1\leq i<j\leq n$\\
then create $n$ new nodes $y_1,\dots,y_n\text{ with }\ml(\la a,y_i \ra)=\{S\}$,
 $\ml(y_i)=\{C\}, \text{ and }y_i\dot\neq y_j\text{ for }1\leq i<j\leq n$

\item 
$\leq$-rule: If 1. $(\leq n S.C)\in \ml(z), z$ is not blocked, and\\
2. $\#S^G(z,C)>n$  and there are two $S$-neighbors $x,y$ of $z$ with
 $C \in \ls(a) \cap \ly$ and not $x \dot\neq y$\\
then 1. If $x$ is a nominal node, then $Merge(y,x)$\\
 2. Else if $y$ is a nominal node or an ancestor of $x$, then $Merge(x,y)$\\
3. Else $Merge(y,x)$ 

\item
$o$-rule If \: for some $o\in N_I$ there are 2 nodes $a,b$ with $o\in\ls(a)\cap\ls(b)$ and not $a\dot\neq b$ \\
then $Merge(a,b)$

\item
NN-rule: If 1.$(\leq n S.C)\in \ls(a), a$ is a nominal node, and there is a blockable 
 $S$-forward-neighbor $b$ of $a$ such that $C\in \ls(b)$ and $a$ is a successor of $b$ \\
2. There  is no $m$ such that $1\leq m\leq n, (\leq m S.C)\in \ls(a)$ 
 there exist $m$ nominal $S$-forward-neighbors $z_1,\dots,z_m$ of $a$ with 
 $C \in \ml(z_i) \text{ and } z_i \dot\neq z_j \text{ for all } 1\leq i < j \leq m$ then\\
1. Guess $m$ with $1\leq m \leq n \text{ and set }\ls(a)=\ls(a)\cup\{(\leq m S.C)\}$ \\
2. Create $m$ new nodes $y_1,\dots,y_m \text{ with } \ml(\la a,y_i \ra)= \{S\},$ 
  $\ml(y_i)=\{C,o_i\}$ for each $o_i\in N_I$ new in \textbf{G}, 
 and $y_i\dot\neq y_j\text{ for }1\leq i<j\leq m$

\end{itemize}

Assume $c\in\vcs$, $q,q_1,q_2$ are markers new to $\ms$. 
For the rules below, by "create a $U=Rg$ (w.l.o.g.) successor of $a$", we mean that we create a successor of $a$ w.r.t $R$ (say $a'$) and then create a $g$ successor of $a'$ (say $c$), such that $\la a,a' \ra=\{R\}$ and $\la a',c \ra=\{g\}$.

\begin{itemize}
\item
\textbf{R$\exists_cU$}: If $(\exists_c U_1,U_2.r)\in \ls(a)$, $a$ is not blocked, and there are no $U_1$ and $U_2$ neighbours of $a$ (say $c_1$ and $c_2$) such that $q_1\in\mss(c_1)$, $q_2\in\mss(c_2)$ and $(q_1 \:r\: q_2)\in\cts$,\\
Then, create a $U_1$ successor of $a$ (say $c_1$) and a $U_2$ successor of $a$ (say $c_2$), and add $(q_1\:r\:q_2)$ to $\cts$, $q_1$ to $\mss(c_1)$ and $q_2$ to $\mss(c_2)$.

\item
\textbf{R$\forall_cU$}: If $(\forall_c U_1,U_2.r)\in \ls(a)$, $a$ is not blocked, there exist $U_1$ and $U_2$ neighbours of $a$ (say $c_1$ and $c_2$), but there exist no markers $q_1$ and $q_2$, such that $q_1\in\mss(c_1)$, $q_2\in\mss(c_2)$ and $(q_1 \:r\: q_2)\in\cts$,\\
Then, add $q_1$ to $\mss(c_1)$, $q_2$ to $\mss(c_2)$ and $(q_1\:r\:q_2)$ to $\cts$.

\textbf{R$\exists_ci$}: If $(\exists_c U_1,\{i\}.r)\in \ls(a)$, $a$ is not blocked, and there is no $U_1$ neighbour of $a$ (say $c_1$) such that $q_1\in\mss(c_1)$, $q_2\in\mss(i)$ and $(q_1 \:r\: q_2)\in\cts$,\\
Then, create a $U_1$ successor of $a$ (say $c_1$) and add $q_1$ to $\mss(c_1)$, $q_2$ to $\mss(i)$ and $(q_1\:r\:q_2)$ to $\cts$.
Analogously for the symmetric case of $(\exists_c \{i\},U_1.r)$.

\item
\textbf{R$\forall_ci$}: If $(\forall_c U_1,\{i\}.r)\in \ls(a)$, $a$ is not blocked, there exists a $U_1$ neighbour of $a$ (say $c_1$), but there exist no markers $q_1$ and $q_2$, such that $q_1\in\mss(c_1)$, $q_2\in\mss(i)$ and $(q_1 \:r\: q_2)\in\cts$,\\
Then, add $q_1$ to $\mss(c_1)$, $q_2$ to $\mss(i)$ and $(q_1\:r\:q_2)$ to $\cts$.
Analogously for the symmetric case of $(\forall_c \{i\},U_1.r)$.

\item
\textbf{R$\neg\exists_cU$}: If $\neg(\exists_c U_1,U_2.r)\in \ls(a)$, $a$ is not blocked, there exist $U_1$ and $U_2$ neighbours of $a$ (say $c_1$ and $c_2$), but there exist no markers $q_1$ and $q_2$, such that $q_1\in\mss(c_1)$, $q_2\in\mss(c_2)$ and $\neg(q_1 \:r\: q_2)\in\cts$,\\
Then, add $q_1$ to $\mss(c_1)$, $q_2$ to $\mss(c_2)$ and $\neg(q_1\:r\:q_2)$ to $\cts$.

\item
\textbf{R$\neg\exists_ci$}: If $\neg(\exists_c U_1,\{i\}.r)\in \ls(a)$, $a$ is not blocked, there exist a $U_1$ neighbours of $a$ (say $c_1$), but there exist no markers $q_1$ and $q_2$, such that $q_1\in\mss(c_1)$, $q_2\in\mss(i)$ and $\neg(q_1 \:r\: q_2)\in\cts$,\\
Then, add $q_1$ to $\mss(c_1)$, $q_2$ to $\mss(i)$ and $\neg(q_1\:r\:q_2)$ to $\cts$.

\item
\textbf{R$\geq_c$}: If $(\geq_cn.g)\in\ls(a)$ and there do not exist $n$ $g$ $successors$ of $a$, $c_1,\dots, c_n$ such that $c_i\dot\neq c_j$ for $1\leq i < j\leq n$, \\
Then create n new concrete nodes $c_1,\dots c_n$ with $\ls(\la a,c_i\ra)=\{g\}$ and $c_i\dot\neq c_j$ for $1\leq i < j\leq n$.

\item
\textbf{R$\leq_c$}: If $(\leq_c n g)\in \ls(a), a$ is not blocked, there exist more than $n$ $g$-successors of $a$, and there are two $g$-successors $c_1,c_2$ of $a$ without $c_1 \dot\neq c_2$,\\
Then, If $c_1\in N_{cI}$, then $Merge(c_2,c_1)$;
else if $c_2\in N_{cI}$, then $Merge(c_1,c_2)$;
else, either $Merge(c_1,c_2)$ or $Merge(c_2,c_1)$.

\item
\textbf{R$connect$}: If $(q_1\:r\:q_2)\in\cts$ and there are nodes $c_1,c_2$ such that $q_1\in\mss(c_1)$, $q_2\in\mss(c_2)$, $Symbol(q_1)=q_1'$ and $Symbol(q_2)=q_2'$,\\
Then add $(c_1\:r\:c_2)$ to $\mn$.

\item
\textbf{R$complete$}: If abstract node $a$ and its descendant $b$, pass the BC-2, then guess completions for $cNet(a)$ and $cNet(b)$.

\end{itemize}

We elaborate on the design of the above algorithm.
While in rules like \textbf{R$\exists_cI$}, we have created the successors in one go, still, we have not connected the concrete nodes in a constraint straight-away.
We do it indirectly by inserting markers, and the completion rule \textbf{R$connect$} adds constraints depending upon markers.
This strategy was required to ensure the case handled by \textbf{R$\neg\exists_cU$} and \textbf{R$\neg\exists_ci$}.

\subsection{Strategy of Rule Application }

The tableau algorithm is initialized as described.
The expansion rules are applied according to the following order of priority:
\begin{itemize}
\item If there is a clash, return unsatisfiable
\item If the completion system is complete, return satisfiable
\item If $Rconnect$ is applicable, it is applied with next highest priority
\item If $Rcomplete$ is applicable, it is applied with next highest priority
\item If the $o$-rule, is applicable, apply with next highest priority
\item If the $\leq$- and the $NN$- rule is applicable, then apply it first to nominal nodes with lower levels(before they are applied to nodes with higher levels). In case they are both applicable to the same node, the $NN$- rule is applied first.
\item All other rules are applied with lower priority
\end{itemize}
It may be noted that the relative ordering of the $\sroiq$ rules is still maintained.
Rules are applied preferably to nodes of lower levels, as explained in $\sroiq$ 

$ $\\
\textit{The KB satisfiablity of a $\sroiqc$ KB is sound, terminating and complete.}
For sketches of the corresponding proofs, the reader is directed to the attached Appendix.

\section{Issues with Extensions}

Thus we have presented a logic which provides constraint modelling constructs to $\sroiq$.
However, we may want to have even more expressive constraint modelling constructs in the logic.
However, we considered several of these expressive constructs and realised that they pose unavoidable, fundamental problems.
These are ultimately related to the patchwork property.

In the current logic, when we have a blocking in $\ms$, while constructing the augmented tableau, we unravel the blocked region into a infinite structure of repeating units.
When we construct the constraint network of the tableau $N$, we do it incrementally, one unit at a time.
Each unit causes the introduction of some new constraints into $N$.
Currently, nodes of these units form constraints with either the nodes of the preceding unit or with a finite set of concrete safe simplex nodes.
These nodes form the common variables between the constraints that are associated with a unit, and the existing network $N$.
Thus, this is a finite set, and we can ensure that the network formed by these common variables is complete, by way of being isomorphic to a network in the completion system ($cNet$).
Thus we can safely apply the patchwork property (among others) to argue that the infinite constraint network associated with these repeating nodes is satisfiable.

\paragraph*{Non simple roles} $ $\\
Consider for example, the restriction that the abstract role used in a path must be a simple role.
Let us see why this causes problems.
If the role is not restricted to be simple, then it can, for eg, be transitive. 
This means that the $R$ successor of some node could be any number of units away from it.
It could happen that a concrete node in one repeating unit forms constraints with the concrete nodes of all the units above it.
Thus we can no longer ensure that the common variables form a complete network, and hence cannot apply the patchwork property.
Thus we can no longer ensure that $N$ is satisfiable.

\paragraph*{Abolish PNF form}$ $\\
Consider another extension. 
Though the PNF form seems to allow us to model many of the constraints that one faces in real domains, it does seem pretty restrictive.
So, lets see what happens if we allow this.
We know that different units can have abstract role edges to nominal tableau nodes (say $R^-$ edges to node $e$).
Now if $\forall_c Rg,Rg.r\in\lt(e)$, then potentially we can again have constraints between the concrete nodes of repeating units arbitrarily far away.
We again can't use the patchwork property and lose out on satisfiability of $N$.

Perhaps a new class of constraint systems may be able to work with these conditions too, but that is left for later analysis.


\chapter{Grounded Circumscription for SROIQ}
\label{chapter3}

\section{A decision procedure for $\gcsr$}

We present here, algorithms to implement Grounded Circumscription for $\sroiqc$.
The algorithms are intuitive and applicable to any decidable logic as long as the logic supports GCIs, inverse and nominal constructs employed in the algorithms. 
Here we use $\sroiqc$ as the base logic to illustrate the algorithms.

A $\gcsr$-KB is the tuple $(K,M)$ where $K$ is the $\sroiqc$ Knowledge Base (KB) and $M$ is the set of minimized predicates.
The minimized predicates may be atomic concepts or roles.
Let $N_I$ be the set of individuals, explicitly named in $K$.
Let $Nom$ be the union of nominal concepts corresponding to the above named individuals.
\begin{align*}
Nom = \{a \mid a\in N_I\}
\end{align*}
For any two models $\mi$ and $\mj$ of $K$, $\mi$ is preferred over $\mj$ w.r.t. $M$ (written $\mi \prec_M \mj$), iff all of the following hold
\begin{enumerate}
\item $\Delta^\mi=\Delta^\mj$ and $a^\mi=a^\mj$ for every $a\in N_I$
\item $W^\mi\subseteq W^\mj$ for every $W\in M$; and
\item There exists a $W\in M$ such that $W^\mi\subset W^\mj$
\end{enumerate}
A model $\mi$ of $K$ is called a grounded model w.r.t. $M$ if all of the following hold :
\begin{enumerate}
\item $C^\mi\subseteq \{b^\mi\mid b\in N_I\}$ for each concept $C\in M$; and
\item $R^\mi\subseteq \{(a^\mi,b^\mi)\mid (a,b)\in N_I\times N_I\}$ for each role $R\in M$
\end{enumerate}
An interpretation $\mi$ is a GC-model of $(K,M)$ if it is a grounded model of $K$ w.r.t. $M$, and $\mi$ is minimal w.r.t. $M$, i.e., there is no model $\mj$ of $K$ with $\mj \prec_M \mi$. 
A GC-$\sroiqc$-KB is said to be GC-satisfiable if it has a GC-model.
A statement (GCI,concept assertion or role assertion) $\alpha$ is a logical consequence(a GC-inference) of $(K,M)$ if every GC-model of $(K,M)$ satisfies $\alpha$. 

We have two Tableau procedures $InitTab$ and $minTab$ based on $Tableau1$ and $Tableau2$ of \cite{lcws}. 
The $InitTab$ computes an initial grounded model of the knowledge base. 
Each iteration of the $minTab$ Tableau tries to produce a model (tableau) which is preferred over the model it was initialized with.
These procedures are used in a coordinated way, as described by the algorithm $modelFinder$, to test the GC-satisfiability of the KB.

\subsection{InitTab} 

We assume that all concepts which are not a part of $M$, are allowed to freely vary.
If it is not so, then we follow the procedure outlined in \cite{lcws} to transform the GC Knowledge Base such that it has no fixed concepts.
We assume Unique Name Assumption to be valid.
We assume the concept $C$ to be atomic. 
If $C$ is not atomic, introduce a GCI $A\equiv C$ and perform the inference procedure w.r.t. $A$.

In the KB, let $M_c$ represent the set of minimized predicates.
Let $M$ consist of $C_i$ (minimized concepts) and $R_j$ (minimized roles), for $1\leq i\leq n$ and $1\leq j \leq m$.

\begin{algorithm}
\DontPrintSemicolon 
\KwIn{$\gcsr$ Knowledge Base $(M_c, \mk)$}
\KwOut{Grounded model $GM$ if it exists}
 For $C_i\in M_c$ assert $C\sqeq Nom$\;
 For $R_j\in M_c$ assert $\exists R_j.\top \sqeq Nom$ \;
 For $R_j\in M_c$ assert $\exists R_j^-.\top \sqeq Nom$ \;
 Run the $\sroiqc$ tableau algorithm\;
 
\eIf{ Clashes occur}
{\Return  Unsatisfiable}
{\Return Grounded Model $GM$}
\caption{{\sc InitTab}}
\label{algo:InitTab}
\end{algorithm}

If InitTab produces a complete and clash-free completion system, then the resulting completion system is a grounded model for the $\gcsr$ KB.

\subsection{MinTab : }

The algorithm is executed on success of $initTab$ .
Let the grounded model produced by $initTab$ be $GM$.

$GM$ is given as input to $minTab$. 
$minTab$ extracts the extensions of minimized predicates from the completion system.
Let the extension of the concept $C_i$ be $C_i^{ext}$ for  $1\leq i\leq n$. 
Let the extension of the role $R_j$ be $R_j^{ext}$ for  $1\leq j\leq m$.
From $R_j^{ext}$ (the extension of the role $R_j$), we extract 
\begin{itemize}
\item
$R_j^{ext,\:dom}$ : the domain for the extension, 
\item
$R_j^{ext,\:range}$ : the range for the extension. 
\end{itemize}

For every $p\in R_j^{ext,\:dom}$, we define $R_j^{ext,\:range,\:p}$ as $\{p'\mid\la p,p'\ra\in R_j^{ext}\}$.

$miniTab$ algorithm involves performing the following for the KB $(\ck,M_c)$ :
\begin{enumerate}
\item Assert $C_i \sqeq C_i^{ext}$ for each concept $C_i$
\item Assert $\exists R^-_j.\top \sqeq R_j^{ext,\:range}$ for each role $R_j$ (LHS represents the set of nodes with incoming $R$ edges)
\item Assert $\exists R_j.\top \sqeq R_j^{ext,\:domain}$ for each role $R_j$ (LHS represents the set of nodes with outgoing $R$ edges)
\item Assert for all role $R_j$ and for each $p\in R_j^{ext,\:domain}$ : $R^-_j.\{p\}\sqeq R_j^{ext,\:range,\:p}$ (LHS represents the set of nodes with incoming $R$ edges, where the edges start from the node $p$)
\item
Activate the preference clash.
\item
Run the $\sroiqc$ tableau algorithm.
\end{enumerate}

We now describe the preference clash.
Let $C^E$ be the extension of a concept $C\in M_c$ extracted from a (possibly incomplete) completion system which is currently being operated upon by completion rules as a part of the $miniTab$ algorithm mentioned above.
Similarly, we extract $R^E$.
A \textit{preference clash} is said to occur if for all $P\in M_c$, $P^{ext}=P^{E}$.
It may be noted that the added clashes are in addition to the ones internal to the tableau algorithms. 
If any clash is encountered, $minTab$ backtracks in attempt to produce a clash-free completion system / model.

If $minTab$ produces a complete and clash-free completion system, then the resulting completion system represents a model preferred over the input model w.r.t. the circumscription pattern.
If $minTab$ produces a complete and clash-free completion system, then this model is fed as an input to $minTab$ again. 
This continues until no more clash-free models can be obtained. 
This final model is a grounded circumscription model (GC-model) of $(\mk,M_c)$.

\subsection{modelFinder algorithm}

This algorithm specifies the interaction between $minTab$ and $InitTab$, and controls their execution.
It receives a $\gcsr$ KB $(\mk,M_c)$ as input, and either returns unsatisfiable or produces a GC-model for the input KB.

First of all, $initTab$ is run.
If it produces a grounded model, then it is fed as input to $minTab$.
If a clash-free model is thereby produced, then the resultant model is fed as an input back into $minTab$.
This continues till no more clash-free models can be produced  by $minTab$.
The resultant model is the GC-model of the input KB.

It may be noted that moth $minTab$ and $initTab$ are non-deterministic algorithms.
So, if required, $modelFinder$ may cause the algorithms to backtrack to an earlier non-deterministic choice point.
As a result the 2 algorithms will strive to produce alternate models.
This may be necessary in order to find a clash-free model if the model output by $minTab$ has clashes (preference clashes or entailment clashes).

\subsection{Other Inference Problems} 

Unlike standard DLs, other inference problems are not reducible to KB satisfiability.
Below we describe the extra clash conditions whose presence in the GC-model, have implications for other inference problems besides KB's GC-satisfiability.
For every inference problem, a specific clash condition is considered.
These clash conditions are called the entailment clash conditions.

\begin{itemize}
\item Instance checking($C(a)$) : The GC-model graph $F$ contains a clash if $F$ has $C\in\ls(a)$.
If this clash is present in the final GC-model produced by $minTab$, then backtracking must ensue.

Thus, $a$ is an instance of concept $C$, if there is no GC-model which is free from this clash.

\item Concept satisfiability($C$) : The GC model graph $F$ contains a clash if $F$ has $C\in \lx$ for any node $x$.

The presence of this clash means that the concept is GC-satisfiable.
Else, backtracking ensues in an attempt to produce another GC-model which may  have this clash.
If no GC-model has this clash condition, then $C$ is not GC-satisfiable.

\item Concept subsumption($C\sqeq D$) : Reduces to concept proving unsatifiability of $C\cap\neg D$
\end{itemize}


\section{Conclusion and Future Work}

We have presented a inference procedure for the expressive DL $\gcsr$. 
It has expressive constraint modelling features and can be used to perform closed world reasoning. 

The correctness of the algorithm is ensured because it adheres to the grounded circumscription framework.
Using the constructs of $\sroiq$, it just provides a new intuitive way, to perform the tasks outlined in the framework.
In this way, it is an implementation of the abstract algorithm whose correctness is elaborated in \cite{lcws}.

$ $\\
However, this does not diminish the usefulness of the algorithm provided here, because a detailed algorithm exists only for $ALCO$.
Though the authors predicted that such algorithm exists for any decidable language, we currently do not have the specification of such an algorithm for $\sroiq$.

Further, the algorithm that did exist for $ALCO$ uses special completion rules for doing so.
However with the provided algorithm, we can achieve grounded circumscription using existing implementations of $\sroiq$ reasoners.

$ $\\
Many tasks still remain, including finding the complexity of the presented logic, optimizing the tableau decision procedure, extending circumscription to prioritized circumscription, amongst others. 
Discovering measures to cope with the non-determinism associated with the inference procedure would greatly ease the practical applications of the logic. 


\appendix
       \chapter{Analysis of $\sroiqc$ tableau algorithm}
\label{appendixA}

Here we present sketches of proofs of correctness for the $\sroiqc$ tableau decision procedure.

\section{Termination}

Termination of the tableau procedure is treated in a manner analogous to the treatment of the tableau procedures in \cite{sroiq} and \cite{shoiq}.
The argument for termination, in case of $\sroiq$ and $\shoiq$, was made on basis of the completion system meeting certain properties, which together, guaranteed termination of the decision procedure.
$\sroiq$ termination was ensured on account of the following properties of the completion system and the completion rules :
\begin{enumerate}
\item
The structure among the blockable nodes is tree shaped i.e. no blockable node has more than one predecessor node.

\item
All rules, except the shrinking/merging rules, add new nodes and edges or append to label sets, but do not remove any nodes or edges and do not remove contents from any label set.

\item
Only generating rules add new nodes, and these rules are triggered at most once for a given concept in the label set of a given node.

\item
If any node $a$ has a blockable successor, then it must have been created by the application of a generating completion rule to the node $a$ i.e. blockable nodes are not inherited (during merging etc.).

\item
The length of a path between a blockable node and its descendant is bounded.

\item
There is a bound on the number of nominal nodes that can be added to the completion system using the $NN$ rule.

\item
The constructs of $\sroiq$, which were not included in $\shoiq$, do not affect the termination of the tableau algorithm.
\end{enumerate}

We propose that the above properties are still met by the abstract parts of the completion system, inspite of new completion rules which manipulate the abstract nodes and edges of the completion system and the new blocking mechanism.
Further, the completion rules that affect concrete nodes and edges, manipulate the completion system in a way that $\ms$ meets certain other properties required to ensure termination.
Lastly, we argue that the remaining completion rules do not affect termination.\\

First, we discuss the properties met by the abstract part of $\ms$ and how they still hold :
\begin{enumerate}
\item
\textit{The depth of the abstract node tree under a nominal node is bounded.}\\

This related directly to the new blocking mechanism.
We must show that the depth can not be increased beyond a certain point because by that point blocking must necessarily happen.

Consider the blocking mechanism, and its checks.
One intuitive way to conceptualize the blocking mechanism is to consider that a node $a$ directly blocks its descendant $b$, if $b$ is a "repetition" of $a$.
The notion of "repetition" is well defined : $b$ is a repetition of $a$, if $a$ and $b$ have the same values for certain properties.

The blocking checks systematically lay down what these properties are.
We argue that after a particular depth from the abstract node $a$, there is guaranteed to be a descendant $b$, which has the same property values as $a$, and thus blocking is guaranteed.
Hence there is a limit on the number of descendant abstract nodes that an abstract node can have, in the completion system.

Next we explain what the properties are and show why their values repeat.\\

\textbf{BC-1} : \\

This check requires all of the following :
\begin{itemize}
\item
both $a$ and $b$ must have equal label sets ($\ls(a)=\ls(b)$), 
\item
their predecessors must have equal label sets ($\ls(a_p)=\ls(b_p)$)
\item
they must have the same kind of edge connecting them to their respective predecessor i.e. $\ls(\la a_p,a \ra)=\ls(\la b_p,b \ra)$.
\end{itemize}

So, the properties that we consider in this BC check are the label sets of the node, its predecessor and the label set of the edge between them.
Repetition in the values of these properties leads to the nodes passing BC-1.
Hence, according to BC-1, a descendant $b$ would be considered as the repetition of a ancestor if it has the same label set as the ancestor, if it's predecessor's label set os the same as that of the ancestor's predecessor and it if the edge joining the node to its predecessor is labelled just as the one joining the ancestor to its predecessor.

We first prove some related propositions ...

\begin{itemize}

\item

\textit{The concepts that form the label set of a blockable node come from a finite, well-defined set.}\\

For a given KB, the concepts that constitute the label set of a \textit{blockable node}, must be from the set $clos(\mt,\mr)$.
A blockable node can, obviously, not contain any nominal concepts, and it therefore, can also not contain the extra concepts that NN-Rule introduces.
All the non internal concepts that the label set of any node can contain come from this finite set.
Also, repetitions of any concept in the label set is not permissible.

So, there is a bound on the number of distinct concepts (concrete or otherwise) that can appear in $\ls(a)$ for any blockable node $a$. 
Let this number be $max_{concepts}$.
So, after having at most $2^{max_{concepts}}$ blockable descendants, if another descendant exists, the it would have to have a label set which would be exactly the same as one of its blockable ancestors.

Thus for a KB, the set of distinct internal constructs that may appear in node's label set is fixed and bounded.

\item

\textit{The label set of an abstract edge is a subset of a finite, well-defined set}\\

The label set of abstract edges is labelled with a subset of the finite set $N_{aR}$.
The power-set of $N_{aR}$ is a finite set, and hence the number of distinct labels of an abstract edge is bounded to a fixed value (say $w_2$).

\end{itemize}

Using the results above, we can establish that we have a bounded number of choices for selecting a distinct/non-repeating label set for the descendant $b$ (say, $w_1$ choices), a fixed number of choices for selecting the label set for the predecessor $b_p$ of $b$ ($w_2$), and a fixed number of choices for selecting the label set for the edge $\la b_p,b \ra$ ($w_1$ again).
Thus, after $w_3=w_1*w_2*w_1$ distinct values for the properties BC-1 tests, there is guaranteed to be a repetition in all the three properties.\\

\textbf{BC-2} :

According to this check $b$ is the repetition of $a$ if they have the same edges joining them to their respective concrete successors, if $b_p$ has this property wrt $a_p$ and if the network formed by $relevantCNodes(b)$ is the same as $relevantCNodes(a)$.

We present some helping propositions first.

\textit{For an abstract node $x$, the number of concrete nodes that can be generated is bounded.}\\

New concrete nodes can be introduced only by the following : $\geq ng$ (corresponding rule fires once for node $x$) and $(\exists_c U_1,U_2.r)$ or $(\exists_c U_1,\{i\}.r)$ (rule fires once for the predecessor of $x$).
Now, the number of such constructs in the KB is limited.
Thus there can be only a fixed number of concrete nodes of any abstract node.
For later analysis let this number be $max_{cNode}$.\\

\textit{For an abstract node, the number of abstract nodes that can be generated as successors is bounded.}\\

For a given abstract node, the no of ancestors is bound to be 1 (on account of the treelike property).
Further, the number of successor abstract nodes is also bounded.
This fact regarding the out-degree of an abstract node, follows from the proofs of $\sroiq$, if we treat $\exists_c U_1,U_2.r$ and $\exists_c U_1,\{i\}.r$ as being equivalent to $\exists R.\top$ in terms of how they manipulate the abstract node structure of the completion system.
These rules fire once for the node in whose label set they are in present (say $x$), and even merging of the newly constructed node back into $x$ does not cause a re-fire.
For later analysis, let this maximum number of abstract nodes be $max_{aNode}$.

$ $\\
We now return to our original task.
The total number of successors for a descendant $b$, and its predecessor $b_p$ is $w_4=2*max_{cNodes}$.
So, we have $w_5=2^{w_4}$ distinct options about the concrete successors that $b$ and $b_p$ can have.
Further let the number of concrete roles in KB be $max_{cRoles}$.
This is a finite number because the set of concrete roles $N_{cR}$ is finite, and the label set of any concrete edge would have to be labelled using roles from this set only.
Correspondingly, we would have $w_6=2^{max_{cRoles}}$ distinct label sets for concrete edges.

Now we can have at most $w_6$ label sets for each of the concrete edges for each of the $w_4$ concrete nodes associated with $b$ and $b_p$.
Thus there are $w_4^{w_6}$ options.

Finally, there can be say $w_7$ constraints between the $w_4$ nodes, each of them capable of being labelled with any one of, say, $w_8-1$ constraint relations.
We can treat the absence of a constraint as another option.
So, the each of the $w_7$ constraints can be labelled in $w_8$ ways, giving us $w_8^{w_7}$ options.

Thus, after $w_9=\: w_4^{w_6}*w_8^{w_7}$ descendants, we will have a repetition in the properties BC-2 tests.
After $w_10=w_9*w_3$ there would be a descendant $b$ which passes BC-2 with $a$.

$ $\\
\textbf{BC-3} :\\

The size of $relativeCNodes(b)$ for any descendant $b$ i finite. Similarly, the nodes constituting $fixedCNodes$ must be the concrete successors of nominal nodes (but not the nominal nodes created by the NN rule).
Clearly the number of such nominal nodes is fixed, and hence the number of concrete nodes in $fixedNodes$ is also fixed.
Thus, the number of nodes in $cNetNodes(b)$ is also fixed, and hence the number of pairs of such nodes is also fixed, say $w_{10}$.

Now any of these pairs can be labelled by any one of say $w_{11}$ constraint relations.
Thus we have $w_{12}=w_{11}^{ w_{10} }$ options.
Thus, after at most $w_{10}*w_{12}$ options, we would have to have a descendant $b$ that passes BC-3.
This suggests that the depth of the abstract tree can not grow beyond a certain limit.

It may also be pointed out that the approximations used here are grossly large approximations of the actual quantity and we have ignored the correlations that do not allow different parameters to take values independently.
Actually, the conditions of BC-1 highly increase the chances for BC-2 and BC-3 to pass, as well.
Thus the upper bound is a very loose one, but does serve to establish that there does exist a bound.

\item
\textit{There is a bound on the number of abstract nodes $\ms$ can have.}\\
We have seen that the depth of an abstract node tree is bounded.
This abstract node tree must be rooted at a nominal.
Further, there are a limited number of nominal nodes in the system which are created during algorithm initialization.
Other than these, nominal nodes can also be created by the $NN$ rule. 
However, results from $\shoiq$ establish that there is a bound on the number of such nodes, and this fact remains unchanged for both $\sroiq$ and $\sroiqc$.
Thus, we can say that the number of all abstract nodes in the completion system is bounded.
\end{enumerate}

Next we discuss some properties pertaining to the novel aspects of $\sroiqc$ completion system :
\begin{enumerate}

\item
\textit{There is a bound on the number of concrete nodes $\ms$ can have.}\\
We established the existence of a bound on the number of abstract nodes in the completion system.
Further, there is a bound on the number of concrete node successors that an abstract node can have.
Using both the above, we can say that there is a bound on the total number of concrete nodes that $\ms$ can have.

\item
\textit{The size of the constraint template set $\mq$ is bounded}\\
Each of the concrete concepts $\exists_c U_1,U_2.r$, $\exists_c U_1,\{i\}.r$, $\forall_c U_1,U_2.r$, $\forall_c U_1,{i}.r$, trigger (at most once) their respective completion rule which adds an entry into $\mq$.
The number of concepts of the above form, in the KB, would be a fixed number, say $x_1$.
We have established that the number of abstract nodes in the completion system is bounded, to say $x_2$.
If all $x_1$ concepts are there in each of the $x_2$ abstract nodes, then also, there would be $x_1*x_2$ i.e. a bounded number of entries of the form $(q_1'\:r\:q_2')$.

\end{enumerate}

Now, in light of the above established properties, we discuss how the number of times each completion rule can be applied is bounded for a given KB.

\begin{itemize}

\item
\textbf{R$\exists_cU$}, \textbf{R$\exists_ci$}, \textbf{R$\neg\exists_cU$}, \textbf{R$\neg\exists_ci$}, \textbf{R$\forall_cU$}, \textbf{R$\forall_ci$} : Each of the first two rules is triggered once, for each instance of their respective trigger concepts in the label set of an abstract node.
In the worst case, the last 4 rules could fire once for every pair of $U_1,U_2$ successors or for every pair of $U_1$ successor and constraint individual $i$.
But since the number of such $U_i$ successors is also fixed, the rules fire a finite number of times.

Since the number of abstract nodes in $\ms$ is bounded, and the number trigger concepts in the label set of an abstract node is bounded, hence, the number of times these rules get triggered is bounded too.

\item
\textbf{R$\exists_ci$}: This rule belongs to class of "generating rules". 
It adds new abstract edges to the completion system.
Clearly there must be a bound on the number of times this rule could be applied.
If it was not so, then the bound on the total number of abstract nodes in the completion system, that we obtained earlier, would not have been possible.

\item
\textbf{R$\forall_ci$}:
Analogous to the case for \textbf{R$\exists_cU$} with the difference that it adds appends to the label sets of its neighbours.

\item
\textbf{R$connect$}: 
This rule introduces a constraint relation between two existing concrete nodes, based on the patterns in $\mq$.
But we already know that there can be at most a fixed, finite number of concrete nodes in the completion system.
Further, there are a finite number of constraint relations.
Thus, it means that this rule can, at most, introduce all possible constraints between all pairs of concrete nodes. Thereafter, it cant fire any more.

\item
\textbf{R$\geq_c$},\textbf{R$\leq_c$}: Analogous to treatment of \textbf{R$\exists_cg$}.

\item
\textbf{R$complete_c$}:
This rule just introduces new constraints between nodes of a fixed size set, once.
Does not affect termination since it applies only when two nodes pass BC-3, which in turn can happen only a finite number of times, given that there is a bound to the depth to which abstract node trees can grow.

\end{itemize}


\section{Soundness}

To establish the soundness of the tableau algorithm, we need to prove the following :\\\\
\textit{If the tableau algorithm returns satisfiable, then the knowledge base $\la\mt,\mr,\ma\ra$ is consistent.}\\
Alternatively,\\
\textit{If the tableau Algorithm produces a complete and clash-free completion system $\ms$, then there exists a valid augmented tableau for the KB.}\\

We begin by describing a procedure to construct an augmented tableau, using the completion system produced by the tableau algorithm.
We then provide a sketch of the proof regarding the validity of the above constructed augmented tableau.

As proposed, an augmented tableau for $\sroiqc$ is the tuple $T_A = \la T, N \ra $ where 
$T=\la \vat,\vct,\ml^T,\eat,\ect\ra$  is a tableau and $N$ is a constraint network with $V_N = \vct$. 
It is easy to argue that the existence of a model of the KB $\mi=(\Delta_\mi,\cdot^\mi, M_\mi)$ is an implication of the existence of an augmented tableau for the KB.
The construction of the augmented tableau may be divided into steps : the construction of $T$, the addition of constraint relations between the concrete nodes of the tableau to construct $N$.

\subsubsection{Procedure for Constructing $T_A$}

\paragraph{Constructing $T$} :\\
$T$ is obtained from the completion system by standard unravelling of the completion graph $\mathcal{G}$\cite{CDR}.
We explore how the completion system can be used to create a possibly infinite tableau for the knowledge base.
The method we employ is an extension of the one used in \cite{shoiq} and \cite{sroiq}.

To uniquely identify every non-nominal node in the unravelled tableau, we use the notion of an "id-path".
An $id$-$path$ is defined as a possibly empty sequence of a pair of nodes of the form : 
$p = ((a_1,b_1 ),( a_2,b_2 ) \dots (a_i,b_i) \dots (a_n,b_n))$,
where,
\begin{itemize}
\item $a_i, b_i \in \vas \cup \vcs$
\item $a_i,b_i$ are never descendants of directly blocked nodes.
\item only $a_1,b_1$ may possibly be from the set of nominal nodes in completion graph $Nom(\ms)$.
Thus, nominal node pairs, if present, may come only as the fist pair of a path.
\item For all i$\geq$1, $b_{i+1}$ is a successor of $a_i$ for an a role $R\in N_R\cup N_{cF}$ i.e. the edge from $a_i$ to $b_{i+1}$ is explicitly present in $\eas\cup\ecs$.
\item
Further, 
\begin{equation*}
a_i \: =
\begin{cases}
\text{Blocker($b_i$),} & \text{if $b_i$ directly blocked}
\\
\text{$b_i$,}& \text{otherwise}
\end{cases}
\end{equation*}
\end{itemize}

For such an id-path $p$, we define $Head(p) = a_1$, $Head'(p) = b_1$, $Tail(p) = a_n$ and $Tail'(p)=b_n$.
In following paras we will use terms like $p,x,y$ to represent paths.

We define the following sets based on the above definitions :
\begin{align*}
Paths(\ms) &= \{ p \mid p \text{ is a id-path} \}\\
Nodes(\ms) &= \{ p \mid p \in Paths(\cs); \\
&Head(p)=Head'(p) \in Nom(\cs)\}\\
\vat &= \{x\mid x\in Nodes(\ms) \text{ and } \\
&Tail(x),Tail'(x) \in \vas \}\\
\vct &= \{y\mid y\in Nodes(\ms) \text{ and } \\
&Tail(y),Tail'(y) \in \vcs \}\\
Nodes_{Nom}&(\ms) = \{e \mid \text{ $e$ of the form } ((o,o)) \text{ for } o\in Nom(\ms) \}\\
Nodes_{NN}&(\ms)=\{x \mid x\in Nodes(\ms); x\neq ((o,o)), \\
&\text{ for any }o\in Nom(\ms)\}
\end{align*}
$Nodes(\ms)$ is the set of all nodes in the augmented tableau, including both concrete and abstract nodes.
$Nodes_{Nom}(\ms)$ is the set of nominal, and $Nodes_{NN}(\ms)$ the set of non-nominal (hence $NN$) abstract nodes in the tableau.\\

We introduce some more definitions :\\
A tableau node $n\in Nodes(\ms)$ is said to be $simplex$ in case for all pairs $(a_i,b_i)$ in $n$, we have $a_i=b_i$.

A node $n\in Nodes(\cs)$ is said to be $associated$ with the tableau nominal node $e$, if $Head(n)=Head'(n)=e$.

$p=(q|(a,a'))$ indicates that the id-path $p$ is formed by appending the tuple $(a,a')$ to the end of $q$.

A node $x$ is the $edge$-$predecessor$ of node $x'$ if $x'=(x \mid (d,d))$ for some $d\in \vas\cup\vcs$.
Predictably, $x'$ is the \textit{edge-successor} of $x$.
A $safe$ $simplex$ node is a simplex node $n$ such that $n$ is the edge successor of nominal tableau node $e$.

Consider a concrete safe simplex nodes such that $y\in\vct$ and it is of the form $((r,r),(i,i))$ where $r\in\vas$ is a nominal node, $i\in\vcs\cap N_{cI}$ is a constraint individual.
For such nodes with id-paths $((r,r),(i,i))$, we use $i$ as an alias for them.
This is essential because in many concepts in the label sets of tableau nodes such as $\exists_c U,\{i\}.r$, a reference is made to the concrete node $i$.

Assume $R\in N_{aR}$ and $g\in N_{cR}$.
We define the remaining tableau structures :
\begin{align*}
\lt(p)& = \ls(Tail(p))\\
\eat(R)& =\\
&\{(x,x') \in Nodes_{NN}(\ms) \times Nodes_{NN}(\ms) \mid \\
&x' =  (x|(a, a')), \text{ and } Tail'(x') \text{ is a } R\:neighbour \\
&\text{  of } Tail(x),\:\:\:OR\\
&x=(x'|(a, a')), \text{ and } Tail'(x) \text{ is a } Inv(R)\:neighbour\\
&\text{  of } Tail(x')\}\:\:\\
\cup\\
&\{(x,e) \in Nodes_{NN}(\ms) \times Nodes_{Nom}(\ms) \mid \\
&Tail(e) \text{ is a } R\:neighbour \text{ of } Tail(x)\}\:\:\\
\cup\\
&\{(e,x) \in Nodes_{Nom}(\ms) \times Nodes_{NN}(\ms) \mid \\
&Tail(x) \text{ is a } R\:neighbour \text{ of } Tail(e)\}\:\:\\
\cup\\
&\{(e,e') \in Nodes_{Nom}(\ms) \times Nodes_{NN}(\ms) \mid \\
&Tail(e') \text{ is a } R\:neighbour \text{ of } Tail(e)\}\\
\ect(g)& =\\
&\{(x,y) \in \vct \times \vct \mid \\
&y =  (x|(c, c)), \text{ and } c=Tail(y) \text{ is a } g\:successor \\
&\text{  of } Tail(x)\}\\
\end{align*}
One may notice that the tableau no longer maintains any treelike property between its non nominal nodes.
For eg. assume $a'$ is the R successor of  $a$, and $R\sqeq R'$. Let $x$ be a tableau node with $Tail(x)=a$. Now there would be a node $x'=(x\mid (a'',a'))$. ($a''$ would be the same as $a'$ depending on whether $a'$ node is blocked or not in $\ms$).
Now, $\la x,x' \ra \in \eat(R)$, $\la x,x' \ra \in \eat(R')$, $\la x',x \ra \in \eat(Inv(R))$ and $\la x',x \ra \in \eat(Inv(R'))$.
Except the complex roles, all other information has been made explicit.
Whatever remains of the treelike structure information is captured in the form of their edge-successor and edge-predecessor re

Also, notice that in the above two definitions, we have used $Tail(x)$ and not $Tail'(x)$, which means that the properties of the tableau node $x\in Nodes_{NN}(\ms)$ with $Tail'(x)=b=$ (some blocked node in $\ms$), are decided by the node which blocked $b$.

\paragraph{Adding the constraints}:\\
Next we explain a way to link the concrete nodes of the tableau by constraint relations.
This would be a little more involved that the above, and we need to introduce some concepts before proceeding further.

An abstract, non-nominal tableau node $h\in\vat$ is termed a $hook$ if $Tail(h)\neq Tail'(h)$.
Further, all nominal nodes of the tableau are defined to be hooks, by default.
The set of all hooks in the tableau is termed $Hooks(\ms)$.
Hook $h'$ is the successor hook of $h$ ($h$ is the predecessor hook of $h'$), if, $h'=hl$, where all $a_i=b_i$ in $l$, except $a_n=Tail(h')$ and $b_n=Tail'(h')$ for which of course $a_n\neq b_n$ (by definition of a hook).
We define :
\begin{align*}
Hooks(\ms,e)=&\{h\mid h\in Hooks(\ms) \text{ and $h$}\\
&\text{is associated with nominal tableau node $e$}\}
\end{align*}
It is possible to define an ordering on the members of the set $Hooks(\ms,e)$ depending upon the increasing distance of the hook from the nominal tableau node $e$.


We attempt to project the entire constraint network $N$ as being the union of smaller, individually satisfiable constraint networks.
Firstly, we decompose $N$ into the following two constraint networks :

\noindent
Formally, the above translates into :
\begin{align*}
N &=N_{simplex} \cup N_{hook}\\
N_{simplex} &= \{ (y\:r\:y')\:\mid \: y,y'\in\vct \text{ are $simplex$ nodes};\\
&\:(Tail(y) \:r\: Tail(y'))\in\mn \}\\
N_{hook} &= \bigcup_{e\in Nodes_{Nom}(\cs)} \: N_e^{hook}
\end{align*}

$N_{hook}$ consist of constraints in which at least one of the concrete nodes is non-simplex.
In these constraints, either both the nodes in the constraint can be non-simplex; else, if only one of them is non-simplex, then the other must be a simplex node.
$N_e^{hook}$ places the extra restriction that the non simplex node must be associated with $e$.
The next few paras are devoted to explaining just which constraints form a part of $N_e^{hook}$.\\

Hooks are used to divide the non-simplex nodes of the augmented tableau, into disjoint sets.
To formalize the above, we define :
\begin{align*}
Path&Fragments(\cs) =\\
&\{p\: \mid \: p \in Paths(\cs);  Head(p),Head'(p)\notin Nom(\cs);\\
&\text{ all pairs $(a_i,b_i)$ in $p$ satisfy $a_i=b_i$};\}\\
Nodes&_{h}=\\
&\{p\mid p\in Nodes(\ms); p=hl; h \text{ is a hook}; \\
&l\in PathFragments(\ms)\}
\end{align*}
It may be noted that the set $Nodes_{h}(\ms)$ is finite, and includes the hook $h$ itself, but not its successor hook.
For purposes of discussions, each of the nodes constituting $Nodes_{h}(\ms)$, is said to be \textit{"in the domain of $h$"}.
One may notice here that all simplex nodes are in the domain of the nominal tableau node that they are respectively associated with.

The set of non-simplex nodes $associated$ with a particular nominal tableau node $e$, can be broken into disjoint sets, each of which would be associated with some non-nominal hook node.
This property is used to decompose the possibly infinite $N_e^{hook}$, into finite constraint networks (say, $N_{h}$), each of which corresponds to a non-nominal hook $h$, such that :
\begin{align*}
N_e^{hook} = \bigcup_{ h\in(Hooks(\ms,e)-\{e\}) } N_h
\end{align*}

We now describe the construction of $N_h$.
Assume $Tail(h)=a$ and $Tail'(h)=b$.
We introduce the following networks :
\begin{align*}
N_h = N^{int}_h \cup N^{pre}_h \cup N^{ss}_h
\end{align*}

$N_h^{int}$ consists of constraints between concrete nodes, both of which fall in the domain of the hook $h$. (int comes from "internal")
It is defined as :
\begin{align*}
N_h^{int}&=\{(y\:r\:y')\mid\: y,y'\in Nodes_h\cap\vct \:;\\
&(Tail(y)\:r\:Tail(y'))\in\mn\}
\end{align*}
In above case, $Tail(y)$ as well as $Tail(y')$ will both point to nodes in $internalCNodes(a,b)$

$N^{ss}_h$ consists of constraints such that one of the concrete nodes is in the domain of the hook $h$, while the other is some safe simplex node.
(ss comes from safe simplex)
\begin{align*}
N_h^{ss} &= \{(y\:r\:y') \mid y\in Nodes_h;\\
&y'\text{ is a safe simplex node}\\
&(\:Tail(y)\:r\:Tail(y')\:)\:\in\mn;\\
& Tail(y')\in fixedCNodes(a,b) \:\}\\
&\cup\text{ the symmetric case}
\end{align*}

$N_h^{pre}$ is a bit more complicated, and to define it we will use the mapping $\theta$ we introduced in BC-3.
Since this is an injective mapping, its inverse is well defined.
We define $\mu=\theta^{-1}$.
Consider a non-nominal hook $h$ and its predecessor hook $p$.
$N^{pre}_h$ consists of constraints between concrete tableau nodes, one of which is in the domain of the hook $h$, while the other is in the domain of the hook predecessor $p$ (hence the term $pre$).

We define :
\begin{align*}
N_h^{pre} &= \{(y\:r\:y') \mid y\in Nodes_h;\: y'\in Nodes_p; \\
& \text{ for }\\
&(\:Tail(c)\:\:r\:\:\mu(Tail(c'))\:)\:\in\mn \}\\
&\cup\text{ the symmetric case}
\end{align*}
Note that in $\ms$, the constraint is not between $c=Tail(y)$ and $c'=Tail(y')$ but between $c$ and $\mu(c')$.

\subsubsection{Validity of $T_A$}

\noindent
Here, we try to prove the following :\\
\textit{The structure created above is a valid augmented tableau for $\la\calr,\ct,\ma\ra$}

It suffices to prove that the structure satisfies all the propositions that must hold in an augmented tableau.
We discuss them one by one ...

\begin{itemize}
\item \textit{(P$_c\:N$Sat)}: $N$ should be satisfiable.

It is in proving this property that the $\omega$ admissibility of the completion system comes into use.
We use the compactness property to pose the possibly infinite $N$ as the union of finite, satisfiable constraint networks.
Since some of these networks may share variables, hence we use the patchwork property to argue the following : since the individual networks are satisfiable and their common variables form complete networks, therefore, $N$ would also be satisfiable.

We of course have to prove that the individual networks are satisfiable and that their intersection networks (network formed by common variables) are indeed complete.
As stated earlier :
\begin{align*}
N&\:=\:N_{simplex} \:\cup\: \bigcup_{e\in Nodes_{Nom}}\:
\bigcup_{h\in Hooks(\ms,e)-\{e\}} N_h
\end{align*}

%

The complete and clash free property of $\ms$ ensures that $N_{simplex}$ is satisfiable.
This is so because it is defined in a way that there is a 1-to-1 mapping from the variables of $N_{simplex}$ to those of $\mn$ (given by $Tail(y)$), and clearly $\mn$ has to be satisfiable to ensure that $\ms$ is clash-free.

For similar reasons, each of the networks $N_h^{int}$, $N_h^{pre}$ and $N_h^{ss}$ are satisfiable, as well.\\

Now we focus attention on the variables common between these networks.
Consider the constraint network $N_e^{hook}$.
It is constructed incrementally, hook by hook.
Assume that, at the time, $N=N_{simplex}$.
In following lines, we attempt to observe how more constraints are added to $N$.

We have discussed that the hooks can be ordered in terms of their depth from the nominal $e_1$.
Consider the first non-nominal hook : $h_1$ i.e. $h_1$ is the successor hook w.r.t $e$.
The constraint network associated with it is $N_{h_1}$ 
There are some variables common between $N_{h_1}$ and $N_{simplex}$.
Now, by def, there are no variables common between $N_{simplex}$ and $N_{h_1}^{int}$.
But there may be some variables common between $N_{simplex}$ and $N_{h_1}^{pre}\cup N_{h_1}^{fixed}$.
These common variables are such that their $Tail$ points to the set $cNetNodes(b)$, where $b$ is blocked by $a$ (also, $Tail(h_1)=a$, $Tail'(h_1)=b$).

More specifically, for $N_{h_1}^{pre}$, the common vars are those which have their $Tail()$ pointing to nodes of $associatedCNodes(b)$.
The must be concrete edge successor nodes of the edge predecessor of the hook $h$.
For $N_{h_1}^{ss}$, the common vars are the safe simplex nodes which have their $Tail()$ pointing to nodes of $fixedCNodes(a,b)$.
These must be the edge successors of nominal tableau nodes.

Clearly, the network these nodes form is ensured to be complete because the completion rules ensure that the network formed by $cNetNodes(b)$ is a complete one.
Using patchwork property we can say that $N=N\cup N_{h_1}$ is satisfiable.
This directly follows since $N$ and $N_{h_1}$ are both satisfiable, and their intersections is a complete, satisfiable network.\\

Next, when we add the constraints of $N_{h_2}$ to $N$, they will again have common variables which can be again mapped to the variables $cNetNodes(b)$.
$N_{h_2}^{pre}$ may have variables common with $N_{h_1}$.
These would be nodes in the domain of $h_1$ such that their tails point to nodes of $associatedCNodes(b)$.
$N_{h_2}^{ss}$ may have variables common with the variables of $N_{simplex}$, such that their tails would point to nodes of $fixedCNodes(a,b)$.

In this way, we can see that $N=N\cup N_{h_2}$ is satisfiable.
This follows because $N$ was satisfiable.
$N_{h_2}$ is a satisfiable constraint network, because it is isomorphic to a satisfiable network, (which is a subset of the satisfiable network $\mn$).
And the intersection of the above networks is complete and satisfiable because the  network formed by their common variables is isomorphic to $cNet(a)$ and $cNet(b)$.

We point out that the above intersection network is indeed complete.
The constraints between the simplex nodes had been introduced as a part of $N_{simplex}$,
The links between the nodes in the domain of $h_1$ must have been introduced as a part of $N_{h_1}^{int}$.
The links between the nodes in domain of $h_1$ and the simplex nodes, would have been introduced as a part of $N_{h_1}^{ss}$.
Thus, $N$ is satisfiable, even after addition of constraints from $N_{h_2}$.

Working this way we can include the constraint networks for other hooks and nominal nodes, and argue that the resultant $N$ would be satisfiable.

\item
$(P_a)$: For all the propositions of the form $P_a$, please refer to the proof of soundness for $\sroiq$. The augmented tableau for $\sroiqc$ shares these propositions with the tableau for $\sroiq$.

\item \textit{(P$_c\:\exists_cU$)}: 
If $(\exists_c U_1,U_2.r)\in \lt(x)$, then there must exist $y_1,y_2\in\vct$ such that $y_1$ is a $U_1$ successor and $y_2$ is a $U_2$ successor of $x$ and $(y_1 \:r\: y_2)\in N$.

Without loss of generality, assume that $U_1=R_1g$ and $U_2=g'$.
Thus, its given that $(\exists_c Rg,g'.r)\in\lt(x)$.

Clearly, $(\exists_c Rg,g'.r)\in\ls(a)$ for $a=Tail(x)$.
Assume $x\in\vat$, and $Tail(x)=a$ and $Tail'(x)=a'$, where $a$ and $a'$ are not necessarily the same (For eg. if $x$ is a hook, then they would be different and $a'$ would be the blocker of $a$ in the completion system).

For the purposes of the following discussion, a node exists in $\ms$ if it is in the set of non blocked nodes of the completion system $\ms$.
If they exist in $\ms$, let $c_1$ be the $U_1$ neighbour of $a$ and $c_2$ the $U_2$ neighbour of $a$, such that $(c_1\:r\:c_2)$.
If $c_1$ and $c_2$ exist, we can also propose the existence of $a_1$, the $R$ neighbour of $a$, such that $c_1$ is the $g$ successor of $a_1$.

$Tail(x)$ ($=a$) can never point to a blocked node ($Tail'(x)$ may, but not $Tail(x)$).
From this, we may infer that $c_2$ will always point to a non blocked node.

However, $a_1$, the successsor of $a$ can have two cases : its either the directly blocked node, or its not blocked at all.
If $a_1$ is blocked then so will be $c_1$.
Thus, the case when $c_1$ doesn't exist in $\ms$ (because of being blocked) is equivalent to the case when $a_1$ is directly blocked.

We now take cases depending upon the nature of $x$, and $c_1$.

\textbf{Case 1}: \textit{$x$ is a simplex node.}

\begin{itemize}

\item
\textbf{Case 1a}: \textit{both $c_1$ and $c_2$ exist ($\in\vcs$ and not blocked)}

We first show that $x$ has a $Rg$ successor $x_1$.
Now there are two cases for $a_1$ :
\begin{itemize}
\item
$a_1$ could be the successor of $a$, such that $R_1'\in\ls(\la a,a_1 \ra)$ for some $R_1'\sqeqs R_1$.
By construction  of the tableau, there will exist an abstract tableau node $x_1$ such that $x_1=(x\mid (a_1,a_1))$.
Also, by construction, $\la x,x_1 \ra \in \eat(R_1'')\cap\eat(R_1)$.
\item
Alternatively, $a$ could be the successor of $a_1$, with $R_1''\in\ls(\la a_1,a \ra)$ for some $R_1''\sqeq Inv(R_1)$.
By construction, there will exist an abstract tableau node $x_1$ such that $x=(x_1 \mid (a,a))$.
Also, by construction, $\la x_1,x \ra \in \eat(R_1'')\cap\eat(Inv(R_1))$, and $\la x,x_1 \ra \in \eat(R_1)$.
\end{itemize}

In comparison, $c_1$ is constrained to be the successor of $a_1$.
That is, $g'\in\ls(\la a_1,c_1 \ra)$ for some $g'\sqeqs g$.
By construction  of the tableau, there will exist an abstract tableau node $y_1$ such that $y_1=(x\mid (c_1,c_1))$.
Also, by construction, $\la x_1,y_1 \ra \in \eat(g')\cap\eat(g)$.

Similarly, we can prove the existence of $y_2$ such that $Tail(y_2)=c_2$.

Since $x$ is simplex and $c_1$ and $c_2$ are both present in $\ms$, therefore we can conclude that $y_1$ and $y_2$ are simplex nodes as well.

The constraint between $y_{U_1}$ and $y_{U_2}$ is included in the constraint network $N_{simplex}$.\\
$(y_{U_1} \:r\: y_{U_2})\in N$ because : $(Tail(y_{U_1}) \:r\: Tail(y_{U_2})) \in \mn$ i.e $(c_{U_1} \:r\: c_{U_2})\in \mn$.

$ $\\
\item
\textbf{Case 1b}: \textit{$c_1$ does not exist}\\

As mentioned earlier, this is possible only if $a_1$ is directly blocked, by (say) $b_1$.
Since $x$ is simplex, $Tail(x)=Tail'(x)=a_1$.
This means that the situation in $\ms$ is such that $b_1$ blocks $a_1$, and $a$ is the predecessor of $a_1$ and $b_1^p$ is the predecessor of $b_1$.
Then, there must exist $x_1=(x\mid(b_1,a_1))$.\\

But the question is, is $c_1$ in $\vcs$ ? it may be blocked but is it there ?
Now, we know that $\ms$ is complete and clash-free (CCF).
$a$ does not have a non-blocked $Rg$ neighbour $c$ such that $(c\:r\:c')$ where $c'$ is its $g$ successor.
But $a$ is not blocked and the $\ms$ is CCF, thus we can conclude that $a_1$ was created by some rule like $R\exists_c$, which would have created the $R$ successor of $a$ ($a_1$) and the $g$ successor of $a_1$ ($c_1$), all in one fell swoop.

Thus we can now conclude that if $\exists_c Rg,g'.r$ is in some non-blocked node of $\ms$, and it does not have any non blocked $Rg$ and $g'$ successors $c,c'$ such that $(c\:r\:c')\in\mn$, then :\\
$c$ is indeed a member of the set $\vcs$ but is blocked. Further, the constraint $(c\:r\:c')$ must be in $\mn$ because $Rconnect$ rule works irrespective of whether the nodes it connects in a constraint are blocked or not.\\

Now we return to our original task.
We have conjectured the existence of the hook $x_1=(x\mid(b_1,a_1))$.
Since $(c_1\:r\:c_2)\in\mn$, by BC-2, we must have a $g'$ successor of $b_1^p$ (say $c_2'$) and a $R$ neighbour of $b_1^p$ (i.e. $b_1$) and a $g$ successor of $b_1$ (say $c_1'$) such that $(c_1'\:r\:c_2')\in\mn$.

Thus, we will have a $y_1\in\vct$ such that $y_1=(x_1\mid(c_1',c_1'))$.
$y_1$ is clearly in the domain of $x_1$.
We already have a $y_2$ such that $y_2=(x\mid(c_2,c_2))$.

The constraint between $y_1$ and $y_2$ would be made as a part of $N_{x_1}^{pre}$.
$(y_{U_1} \:r\: y_{U_2})$ must be in $N$, because there exists in $\mn$ : $( Tail(y_1) \:r\: \theta^-(Tail(y_2))  )$ = $(c_1' \:r\: \theta^-(c_2) )$ = $(c_1'\:r\:c_2')$.\\

\end{itemize}


\textbf{Case 2}: \textit{$x$ is not a simplex node}\\

Let $x$ be associated with the tableau nominal $e$, and be in the domain of the non-nominal hook $h$.
The case demands that $Tail(x)\in internalCNodes(b_1,b_2)$, where $b_1$ directly blocks $b_2$ in $\ms$.
We can conclude that $c_2$ will always be in $internalCNodes(b_1,b_2)$.

For the $Rg$ successor ($c_1$), there are some options regarding where it can exist in $\ms$.
It may be in $internalCNodes(b_1,b_2)$, $fixedCNodes(b_1,b_2)$, $relativeCNodes(b_1,b_2)$ or it may be blocked.
We have to take cases for all of these cases.

\begin{itemize}

\item
\textbf{Case 2a}: \textit{$c_1 \:\in\: internalCNodes(b_1,b_2)$}

Arguing as in Case 1a, we can propose the existence of a $U_1$ successor of $x$ ($y_1$) and a $U_2$ successor of $x$ ($y_2$), such that $Tail(y_1)=c_1$ and $Tail(y_2)=c_2$.
Furthermore, $y_1$ and $y_2$ can both be argued to be in the domain of the hook $h$.

The constraint between them would be made as a part of $N_{h}^{int}$ i.e.
$(y_1 \:r\: y_2)\in N$ because $(Tail(y_1) \:r\: Tail(y_2))\in \mn$  $\implies$ $(c_1 \:r\: c_2)\in \mn$.

$ $\\
\item
\textbf{Case 2b}: \textit{$c_1 \:\in\: fixedCNodes(b_1,b_2)$}

We can argue the existence of the $g'$ successor of $x$ ($x_2$).
$y_2$ would be in the domain of $h$.

Lets consider the existence of $y_1$.
$c_1\in fixedCNodes(a,b)$ means that there exists a nominal node which is the $R$ successor of $a$, (say $r$), and it has $g$ successor $c_1$ such that $(c_1\:r\:c_2)\in\mn$.
Using the above fact we can establish that $\la x,e \ra \in \eat(R)$, where $e=(r,r)$ is a nominal tableau node.
We can further argue about $y_1$ such that $y_1=(e\mid(c_1,c_1))$.
Clearly, $y_1$ is a safe simplex node.

The constraint between $y_1$ and $y_2$ them would be made as a part of $N_{ss}$.
$(y_1 \:r\: y_2)\in N$ because $(Tail(y_1) \:r\: Tail(y_2))\in \mn$  $\implies$ $(c_1 \:r\: c_2)\in \mn$.

$ $\\
\item
\textbf{Case 2c}: \textit{$c_1\in relativeCNodes(b_1,b_2)$}

Consider the completion system.
The case suggests that $c_1$, the $Rg$ neighbour of $a$ is in $relativeCNodes(b_1,b_2)$.
By definition of $relativeCNodes(b_1,b_2)$, this means that $c_1$ is the $g$ successor of $b_1^p$ (the predecessor of $b_1$).
This means $a$ is a blocker node (i.e. $b_1=a$).

On account of BC-2, one can also argue the existence of some $g$ successor of $b_2^p$ (say $c_1'$, such that $c_1'=\phi(c_1)$) and some $g'$ successor of $b_2$ (say $c_2'$, such that $c_2'=\phi(c_2)$) such that $(c_1'\:r\:c_2')\in\mn$.

Now, consider the tableau.
We can argue the existence of the $g'$ successor of $x$ ($y_2$), which would be in the domain of $h$.
Its also known that $x$ is not simplex, and $Tail(x)$ points to $a$, hence it must be a hook, with $Tail(x)=b_1=a$ and $Tail'(x)=b_2$ (for this case $a=b_1$).
Now there exists a node $x_1$ (the edge predecessor of $x$) such that $x=(x_1\mid(b_1,b_2))$, and $Tail(x_1)=Tail'(x_1)=b_2^p$.
Further, we define $y_1$ as $y_1=(x_1\mid(c_1',c_1'))$.
We can see that $y_1$ and $x_1$ belong in the domain of the predecessor hook (say $h_p$).

Now, the connection between $y_2$, $g'$ successor of $x$, and $y_1$, the $Rg$ successor of $x$ is made as a part of $N_h^{pre}$, where $y_1$ is in the domain of $h_p$ while $y_2$ is in the domain of $h$.
$(y_1 \:r\: y_2)$ must in $N$, since there exists in $\mn$, the following constraint : $(  \theta^-(Tail(y_1)) \:r\: Tail(y_2)  )$ = $(\theta^-(c_1')\:r\:c_2)$ = $(c_1)$

$ $\\
\item
\textbf{Case 2d}: \textit{$c_1$ is blocked}

Consider $\ms$.
We know that a tableau node cannot have its $Tail()$ point to any node that is blocked.
Given that $Tail(x)$ points to $a$, $a$ cannot be blocked.
However, the premise of the case suggests that $c_1$, the $Rg$ successor of $a$, is blocked.
This can happen only in case $a_1$ is blocked.
However, $c_2$ is not blocked and we do have $(c_1\:r\:c_2)\in\mn$.

With regards to blocking the situation is that $b_1$ blocks $a_1$. their respective predecessors are $b_1^p$ and $a$.

$ $\\
Consider now the Tableau.
Using BC-2 we can propose the existence of the $g$ successor of $b_1$ (say $c_1'=\phi^-(c_1)$), and the $g'$ successor of $b_1^p$ (say $c_2'=c_2$) such that $(c_1'\:r\:c_2')\in\mn$.

We can propose the existence of $y_2$ such that $y_2=(x\mid (c_2,c_2))$.
Further there exists $x_1$ such that $x_1=(x\mid(b_1,a_1))$, and $y_1=(x_1\mid(c_1',c_1'))$.
As we can see, $x_1$ is a hook, and $y_2$ is in the domain of $x_1$, while $y_1$ is in the domain of $h$, same as $x$.
The link between $y_1$ and $y_2$ is introduced as a part of $N_{x_1}^{pre}$.
$(y_1\:r\:y_2)\in N$ because there exists in $\mn$ the following relation : $(Tail(y_1)\:r\:\theta^-(Tail(y_2)))$ i.e. $(c_1'\:r\:\theta^-(c_2))$ i.e. $(c_1'\:r\:c_2')$.

\end{itemize}


\item \textit{(P$_c\:\forall_cU$)}: 
If $(\forall_c U_1,U_2.r)\in \lt(x)$, then for all $y_1,y_2$ such that $y_1$ is a $U_1$ successor, $y_2$ is a $U_2$ successor of $s$ and $(y_1\:r\:y_2)\in N$.\\

Without loss of generality, assume that $U_1=R_1g$ and $U_2=g'$.
Thus, its given that $(\forall_c Rg,g'.r)\in\lt(x)$.

Clearly, $(\forall_c Rg,g'.r)\in\ls(a)$ for $a=Tail(x)$.
Assume $Tail(x)=a$ and $Tail'(x)=a'$, where $a$ and $a'$ are not necessarily the same (For eg. if $x$ is a hook, then they would be different and $a'$ would be the blocker of $a$ in the completion system).

Let $x$ have any number of $Rg$ and $g'$ successors in the tableau.
We take pairs of them one by one.
Consider any such $Rg$ successor $y_1$ and $g'$ successor $y_2$.
We have to prove that $(y_1\:r\:y_2)\in N$.
Let $Tail(y_1)$ be $c_1$ and $Tail(y_2)$ be $c_2$.
Further since $y_1$ is the $Rg$ successor, then there must be a $R$ successor of $x$ (say $x_1$) such that $Y_1$ is the $g$ successor of $x_1$.
Let the $Tail(x_1)$ be $a_1$.

We take cases depending upon the nature of $x$ and the location of the successors.
We point out that in any case $y_1$ would be in the domain of the same hook as $x$.
We can always propose the existence of $c_2$ such that $c_2$ is the successor of $a$ and $y_2=(x\mid(c_2,c_2))$, where $a=Tail(x)$ (not necessarily $Tail'(x)$).
So, we only need to worry about $y_1$.

\begin{itemize}

\item
\textbf{Case 1} : $x$ is a simplex node.\\

Let $x$ be in the domain of the nominal tableau node $e$.
Let $Tail(x)=Tail'(x)=a$.\\

\begin{itemize}

\item
\textbf{Case 1a} : $y_1$ is a simplex node i.e. its in the domain of $e$.\\

Its presumed that $x$ has a $Rg$ successor.
Let the $R$ successor be $x_1$.
Now, if $x_1$ is the edge-successor of $x$, then $Tail(x_1)$ is the $R$ successor of $x$, and $R'\in\ls(\la Tail(x),Tail(x_1) \ra)$ for some $R'\sqeqs R$.
Else, $R'\in\ls(\la Tail(x_1),Tail(x) \ra)$ for some $R'\sqeqs Inv(R)$.
Further since $y_1$ must be the $g'$ successor (edge successor) of $x_1$, we have $g'\sqeqs\ls(\la Tail(x_1),Tail(y_1) \ra)$ for some $g'\in g$.

Now, $Y_1$ is bound to be such that $y_1=(x_1\mid(c_1,c_1))$.
This case implies the presence of a $Rg$ neighbour of $a$ (say $c_1$) such that its not blocked, and $Tail(y_1)=Tail(y_1)=c_1$.

Now, we have conjectured the presence of $a$, its $Rg$ successor $c_1$ and $g'$ successor $c_2$.
Clearly, CCF condition of $\ms$ ensures $(c_1\:r\:c_2)\in \mn$.
Because of this, $(y_1\:r:y_2)$ form a part of $N_{simplex}$.

$ $\\
\item
\textbf{Case 1b} : $y_1$ is not a simplex node i.e. its in the domain of some successor hook $h$.\\

$x_1$ cannot be the edge predecessor of $x$ in this case because then $y_1$ would be simplex too, which is contrary to the supposition of the case.
Thus $x_1$ must be the edge successor.

Further $y_1$ can be in the domain of another hook only if its edge predecessor $x_1$ is in the domain of another hook.
Now, $x$ is simplex. 
$y_1$ is in the domain of its successor hook $h$, and so must be its edge predecessor $x_1$.
Thus we conclude that $x_1$ is a hook.
Let $Tail(x_1)=b_1$ and $Tail'(x_1)=b_2$ where $b_1$ blocks $b_2$.

In terms of the blocking scenario, we can conclude that $b_1$ blocks $b_2$, $b_1^p$ is the predecessor of $b_1$ and $a$ is the predecessor of $b_2$.

Now, $Tail(x)=Tail'(x)=a$, and $(\forall_cRg,g'.r) \in \ls(a)$.
We know that $a$ has the $g'$ successor $c_2=Tail(y_2)$.
Further, it has the $R$ neighbour $b_2=Tail'(x_1)$.
Also, $b_1=Tail(x_1)$ has the $g$ successor $c_1=Tail(y_1)$.

By BC-2, this directly means that $b_2$ must also have a $g$ successor $c_1^{down}=\phi(c_1)$.
Also, there must be a $g'$ successor of $b_1^p$, say $c_2^{up}=\phi^-(c_2)$.

CCF condition ( the $Rcomplete$ ) rule ensure that $(c_1^{down}\:r\:c_2)\in\mn$.
By BC-2, using the above, we conclude that $(c_1\:r\:c_2^{up})\in\mn$.

Finally, $(y_1\:r\:y_2)\in N$ would be as a part of $N_h^{pre}$, where $y_1$ is in the domain of $h$ and $y_2$ is in the domain of $e$ (same as $x$).
$(y_1\:r\:y_2)\in N$ because $(Tail(y_1)\:r\:\theta^-(Tail(y_2)))\in \mn$ = $(c_1\:r\:\theta^-(c_2))$ = $(c_1\:r\:c_2^{up})\in\mn$.

\end{itemize}


\item
\textbf{Case 2} : $x$ is not a simplex node.\\

Let $x$ be in the domain of hook $h$.
Clearly, $y_2$ is also in the domain of $h$.
We now take cases about $y_1$.

\begin{itemize}

\item
\textbf{Case 2a} : $y_1$ is in the domain of $h$.\\

This is a simple case and can be argued just like 1a.
$ $\\

\item
\textbf{Case 2b} : $y_1$ is in the domain of a predecessor hook $h_p$.\\

$x_1$ cannot be the edge successor of $x$ in this case, because then $y_1$ would have to be in domain of $h$, which is contrary to the supposition of the case.
Thus $x_1$ must be the edge predecessor.

Further $y_1$ can be in the domain of another hook only if its edge successor $x_1$ is in the domain of another hook.
Now, $x_1$ is in the domain of $h_p$ and its edge successor $x$ is in the domain of $h$.
This is possible only if $x$ is a hook.
Let $Tail(x)=a$ and $Tail'(x)=a'$ where $a$ blocks $a'$.
So, $x=(x_1\mid(a,a'))$.

In terms of the blocking scenario, we can conclude that $a$ blocks $a'$, $a^p$ is the predecessor of $a$ and $a_1$ is the predecessor of $a'$.

Now, $Tail(x)=a$, $Tail'(x)=a'$, and $(\forall_cRg,g'.r) \in \ls(a)=\ls(a')$.
Further, $a_1$, the predecessor of $a'$ is the $R$ neighbour of $a$.
BC-1 ensures $a^p$ is the $R$ neighbour of $a$.

We know that $a$ has the $g'$ successor $c_2=Tail(y_2)$.
We also know that $a_1$ has the $g$ successor $c_2=Tail(y_2)$.
By BC-2, this directly means that $a_p$ must also have a $g$ successor $c_1^{up}=\phi^-(c_1)$.
Also, there must be a $g'$ successor of $a'$, say $c_2^{down}=\phi(c_2)$.

BC-2 and CCF condition ( the $Rcomplete$ ) rule ensure that $(c_1\:r\:c_2^{down})\in\mn$ and $(c_1^{up}\:r\:c_2)\in\mn$.

Finally, $(y_1\:r\:y_2)\in N$ would be as a part of $N_h^{pre}$, where $y_1$ is in the domain of $h_p$ (predecessor hook of $h$) and $y_2$ is in the domain of $h$ (same as $x$).
$(y_1\:r\:y_2)\in N$ because $(\theta^-(Tail(y_1))\:r\:Tail(y_2))\in \mn$ = $(\theta^-(c_1)\:r\:c_2)$ = $(c_1^{up}\:r\:c_2)\in\mn$.

$ $\\
\item
\textbf{Case 2c} : $y_1$ is in the domain of a successor hook $h_s$.\\

This is much the same case as 1b.
The only difference is that $x$ was simplex there (in domain of some nominal tableau node $e$), but here it is in the domain of some hook $h$.

\item
\textbf{Case 2d} : $y_1$ is a safe simplex node.\\

This means that $x$, $y_2$ is in the domain of some hook $h$.
Further, $x_1$ must be a nominal tableau node, and $a_1$ a nominal node in the completion system.

$(\forall_c Rg,g')\in a$, $a_1$ is the $R$ neighbour of $a$, $c_1$ is the $g$ successor of $a_1$, $c_2$ is the $g'$ successor of $a$.
The CCF condition ensures that $(c_1\:r\:c_2)\in\mn$.

Thus $(y_1\:r\:y_2)$ is a part of $N_h^{ss}$.
$(y_1\:r\:y_2)\in N$ because $(Tail(y_1)\:r\:Tail(y_2))$ = $(c_1\:r\:c_2)$ is in $\mn$.

\end{itemize}

\end{itemize}


\item \textit{(P$_c \:\neg\exists_cU$)}: $ $\\
The overall logic would be to show that if $\neg(\exists_c U_1,U_2.r)\in\lt(x)$ and $x$ has some $U_1,U_2$ successors $Y_1,y_2$ such that $(y_1\:r\:y_2)\in N$, then using cases just like in above propositions, we would be able to identify $U_1,U_2$ neighbours of $a=Tail(x)$ (say $c_1,c_2$) such that $(c_1\:r\:c_2)$ would have to be in $\mn$.
But since this would have lead to a clash and $\ms$ is CCF, therfore we prove by contradiction that none of the $U_1$ and $U_2$ successors of $x$ are related by the constraint $r$.


\item \textit{(P$_c \:\leq_c$)}: If $(\leq n g)\in \lt(x)$, then 
$\#\{y_i \mid y_i\in \vct, \la x, y_i\ra\in\ect(g)\}\leq n$.

Let $a=Tail(x)$.
$a$ can not be a blocked node ($Tail'(x)$ may but not $Tail(x)$).
Clearly, $(\leq n g)\in \ls(a)$.
Now, by the CCF condition of $\ms$, there are $\leq n$ $g$ successors of $a$ (say $c_i$).
If there were any more successors than $n$, then it would be a clash.
Correspondingly we will have $\leq n$ edge-successor of $x$ such that $y_i=(x\mid(c_i,c_i))$.
All the $y_i$ are clearly the $g$ successors of $x$.

\item \textit{(P$_c \:\geq_c$)}: If $(\geq n.g)\in \lst$, then  
$\#\{c_i \mid c_i\in \vct, \la s, c_i\ra\in\ect(g)\}\geq n$.

In the completion system, $(\geq n.g)\in\ls(a)$, and by the CCF condition, we have $n$ distinct $g$-successors of $a$ ($c_i$).
Corresponding to every one of those $c_i$, we will have a concrete successor of $x$, say $y_i$ such that $Tail(y_i)=c_i$.
Thus, $x$ also has greater than n distinct $g$ successors.

\item \textit{(P$_c\:\forall_ci$)} and \textit{(P$_c\:\exists_ci$)}: 

This can be treated in a manner very similar to (P$_c\:\exists_c$) and (P$_c\:\forall_c$).
wlog, assume $U_1=Rg$.
The only difference would be that in this case the constraint would always be introduced as a part of $N_h^{ss}$, where $h$ is the hook in whose domain $y_1$, the $Rg$ successor of $x$, is.

\end{itemize}


\section{Completeness}

\textit{If the KB is satisfiable i.e. there exists an augmented tableau $T_a$ for it, then the tableau algorithm returns a complete and clash free completion system.}

Let $T_a=(T,N),\:T=(\vat,\vct,\lt,\eat,\ect)$ for it. We need to prove that the Tableau rules can be applied in a way to yield a complete and clash-free completion system, so that the tableau algorithm returns satisfiable. 
To this end, we use the augmented tableau $T_a$ to "guide" the application of tableau rules.

We introduce the notion of $T_a$ \textit{compatibility} of a completion system $\ms$. 
We define mappings :\\
$\pi_a:\vas\rar\vat$ and $\pi_c:\vcs\rar\vct$. 
By means of these mappings, every node in the $\cs$ is mapped to a node in $T_a$. 
A completion system is said to be $T_a$ $compatible$ to a augmented tableau $T_a$ if it obeys the following propositions. 
For the following we assume: $a,b\in\vas;\:c_1,c_2\in\vcs$; $g\in N_{cF},R\in N_{aR}$.
\begin{itemize}
\item
$(C_a)$: For $C\in clos'(\ct,\calr)$; if $C\in\ls(a)$, then $C\in\lt(\pi_a(a))$
\item
$(C_b)$: If $b$ is $R$-forward-neighbour of $a$, in $\cs$, then $(\pi_a(a),\pi_a(b)\in\eat(R)$
\item
$(C_c)$: If $a\dot\neq b$, then $\pi_a(a)\neq\pi_a(b)$
\item
$(C_d)$: If $\leq nS.C\in\ls(a)$ and $\leq nS.C\notin clos'(\ct,\calr)$, the $\#S^T(a,C)=n$\\

New clauses for $\sroiqc$ :

\item
$(C_e)$: If $c$ is $g$-successor of $a$, in $\ms$, then $(\pi_a(a),\pi_c(c)\in\ect(g)$
\item
$(C_f)$: If $(c_1\:r\:c_2)\in\cn$, then $(\pi_c(c_1)\:r\:\pi_c(c_2))\in N$

\end{itemize}

\textit{A $T_a$ compatible completion graph is clash-free.}

We prove how these condition ensure that the clashes new to $\sroiqc$ e.r.t $\sroiq$ do not occur.

\begin{itemize}

\item
Clash 1:\\
This follows directly from $C_e$ and $C_c$.
If suppose that the clash condition did happen in $\ms$ for node $a$, then it would mean that the $P\leq$ tableau proposition is violated for $\pi_a(a)$.
But this can't happen because the augmented tableau is known to be a valid one.

\item
Clash 2:\\
Assume $neq(q_1\:r\:q_2)\in\cts$, and there exist $c_1,c_2$ such that $q_1\in\mss(c_1)$ and $q_2\in\mss(c_2)$.
Now these could have got into $\mss$ of $c_1$ and $c_2$ only if there was a node $a$ such that $\neg(\exists_cU_1,U_2.r)\in\ls(a)$ and $c_1$ is its $U_1$ neighbour and $c_2$ is its $U_2$ neighbour.
(The case could also be $\neg(\exists_c U,\{i\}.r)$ or $\neg(\exists_c \{i\},U.r)$ but they can be dealt analogously )
Now the clash would occur if $(c_1\:r\:c_2)\in\mn$.

But this would mean that there exists $\pi_a(a)$ such that $\neg(\exists_cU_1,U_2.r)\in\lt(\pi_a(a))$, there exist $U_1,U_2$ successors of $y_1,y_2$ and $(y_1\:r\:y_2)\in N$.
But this would make the augmented tableau invalid since it would not satisfy the proposition $P_c\neg\exists_cU$.
But its given that tableau is, indeed, valid.
Thus, this clash cant happen.

\item
Clash 3:\\
we prove this by contradiction.
Suppose that $\mn$ was, in fact, not satisfiable.
Then we would have nodes in tableau given by the mapping $\pi_c$ such that the constraint network between them is not satisfiable.
But this is not possible, since the augmented tableau is by definition satisfiable.

\end{itemize}

\textit{Applicable rules can be applied to a $T_a\:\:compatible$ system in such a way as to yield a new $T_a\:\:compatible$ system.}

We show this for the new Tableau Rules. For the other rules, the reader may refer to \cite{Horrocks07}.
Assume w.l.o.g., $U_1=g_1,U_2=Rg_2$ for the concrete constructs; $a\in\vas$.

Since $T_a$ is satisfiable, we guess a completion of it, $N^c$.

\begin{itemize}

\item
$(R\exists_cU)$\\
$\exists_c U_1,U_2.r\in\ls(a)$. 
Let $U_1=Rg$, $U_2=g'$.
Then by $C_a$, $\exists U_1,U_2.r\in\lt(\pi_a(a))$.
Since $T_a$ is valid, by $P\exists_cU$ proposition, there must exist a $g'$ successor of $\pi_a(a)$ (say, $y_2$) a $R$ successor of $\pi_a(a)$ (say $x_1$) and a $g$ successor of $x_1$ (say $y_1$), such that $(x_1\:r\:x_2)\in N$.
After application of rule, we create node $a_1,c_1,c_2$ such that $a_1$ is a $R$ successor of $a$, $c_1$ is the $g$ successor of $a$ and $c_2$ is a $g'$ successor of $a_1$.\\
We set $\pi_a(a_1)=x_1;\pi_c(c_1)=y_1;\pi_c(c_2)=y_2$. 
Clearly, the new system is still $T_a$ compatible.

\item
$(R\forall_cU)$\\
This is a deterministic rule.
$\forall_c U_1,U_2.r\in\ls(a)$. 
Let $U_1=Rg$, $U_2=g'$.
Then by $C_a$, $\exists U_1,U_2.r\in\lt(\pi_a(a))$.
Since $T_a$ is valid, by $P\forall_cU$ proposition, all the $U_1$ and $U_2$ successors are related by $r$ constraint.
Consider any such $U_1, U_2$ successors $y_1,y_2$ such that $y_1=\pi_c(c_1)$ and $y_2=\pi_c(c_2)$.
Clearly, $c_1,c_2$ would have to be the $U_1,U_2$ neighbours of $a$, where $\pi_a(a)=x$.
Then by adding $(c_1\:r\:c_2)$ to $\mn$, we can not have any clash.

\item
$(Rcomplete)$: \\
Clearly the involved nodes have their respective counterparts in tableau, given by the mapping $\pi$.
The network in tableau is guaranteed to be satisfiable.
We use the corresponding constraints in $N^c$ to guess completions for this rule.

\item
$(R\leq_c)$: \\
If $(\leq_c n.g)$ is in $\ls(a)$, then it is there in $\lt(\pi_a(a))$ as well.
By $P_c\leq_c$ proposition, we can be sure that $\pi_a(a)$ does indeed have less than $n$ $g$ successors $y_i$.
Now even if $a$ has more than $n$ successors $c_i$, they would still have to point to these $y_i$.
Thus, we merge the required nodes such that ultimately there are less than $n$ $c_i$ that point to these $y_i$, in a on-to-one manner.

\item
The rest of the propositions can be handled in manners analogous to above.

\end{itemize}

Clearly, the initial graph is $T_a$ \textit{compatible}, with the mapping $\pi_c$ empty and $\pi_a$ mapping the nominal nodes and constraint individuals to their images in the augmented tableau. 
Since any applicable rule can be applied in a way to give a $T_a$ compatible system, we continue doing so until no more rules are applicable. 
Thus the completion system at the end would be complete and clash free.


\renewcommand{\baselinestretch}{1.3}

\begin{small}
\addcontentsline{toc}{chapter}{References}
\renewcommand{\bibname}{References}
\bibliographystyle{IEEEtran}
{\normalsize
\bibliography{references} 
}
\end{small}

\end{document}